\documentclass[twocolumn]{aastex63}
\usepackage{CJK}
\usepackage{amsmath}
\usepackage{verbatim}

\usepackage{hyperref}
\usepackage{multirow}

\definecolor{blue}{RGB}{50, 80, 255}

\shorttitle{Infrared Spectroscopy of White Dwarfs}
\shortauthors{Owens et al.}

\begin{document}
\title{Disk or Companion: Characterizing Excess Infrared Flux in Seven White Dwarf Systems with Near-Infrared Spectroscopy}

\correspondingauthor{Dylan Owens} 
\email{dylowens6@gmail.com}

\author[0000-0002-6397-6719]{Dylan Owens}
\affil{Gemini Observatory/NSF's NOIRLab, 670 N. A'ohoku Place, Hilo, Hawaii, 96720, USA}

\author[0000-0002-8808-4282]{Siyi Xu \begin{CJK*}{UTF8}{gbsn}(许\CJKfamily{bsmi}偲\CJKfamily{gbsn}艺)\end{CJK*}}
\affil{Gemini Observatory/NSF's NOIRLab, 670 N. A'ohoku Place, Hilo, Hawaii, 96720, USA}

\author[0000-0003-0192-6887]{Elena Manjavacas}
\affil{AURA for the European Space Agency (ESA), ESA Office, Space Telescope Science Institute, 3700 San Martin Drive, Baltimore, MD 21218 USA}

\author[0000-0002-3681-2989]{S. K. Leggett}
\affil{Gemini Observatory/NSF's NOIRLab, 670 N. A'ohoku Place, Hilo, Hawaii, 96720, USA}

\author[0000-0003-2478-0120]{S. L. Casewell}
\affil{School of Physics and Astronomy, University of Leicester, University Road, Leicester LE1 7RH, UK}

\author[0000-0003-2852-268X]{Erik Dennihy}
\affil{Rubin Observatory Project Office, 950 N. Cherry Ave., Tucson, AZ 85719, USA}

\author[0000-0003-4609-4500]{Patrick Dufour}
\affil{Institut de Recherche sur les Exoplan\`etes (iREx), Universit\'e de Montr\'eal, Montr\'eal, QC H3C 3J7, Canada}
\affil{D\'epartement de physique, Universit\'e de Montr\'eal, Montr\'eal, QC H3C 3J7, Canada}

\author[0000-0001-5854-675X]{Beth L. Klein}
\affil{Department of Physics and Astronomy, University of California, Los Angeles, CA 90095-1562, USA}

\author[0000-0002-4037-3114]{Sherry Yeh}
\affil{W. M. Keck Observatory, Waimea, HI, USA}

\author{B. Zuckerman}
\affil{Department of Physics and Astronomy, University of California, Los Angeles, CA 90095-1562, USA}

\begin{abstract}

Excess infrared flux from white dwarf stars is likely to arise from a dusty debris disk or a cool companion. In this work, we present near-infrared spectroscopic observations with \textit{Keck/MOSFIRE}, \textit{Gemini/GNIRS}, and \textit{Gemini/Flamingos-2} of seven white dwarfs with infrared excesses identified in previous studies. We confirmed the presence of dust disks around four white dwarfs (Gaia~J0611--6931, Gaia~J0006+2858, Gaia~J2100+2122, and WD~0145+234) as well as two new white dwarf brown dwarf pairs (Gaia~J0052+4505 and Gaia~J0603+4518). In three of the dust disk systems, we detected for the first time near-infrared metal emissions (Mg I, Fe I, and Si I) from a gaseous component of the disk. We developed a new Markov Chain Monte Carlo framework to constrain the geometric properties of each dust disk. In three systems, the dust disk and the gas disk appear to coincide spatially. For the two brown dwarf white dwarf pairs, we identified broad molecular absorption features typically seen in L dwarfs. The origin of the infrared excess around Gaia~J0723+6301 remains a mystery. Our study underlines how near-infrared spectroscopy can be used to determine sources of infrared excess around white dwarfs, which has now been detected in hundreds of systems photometrically.

\end{abstract}


\section{Introduction}

The vast majority of stars within the Galaxy will one day reach the last stage of their evolution as white dwarfs. Some main sequence stars are in binary systems with substellar companions, which can be difficult to detect and characterize given the much brighter primary. When the primary reaches the white dwarf stage, it becomes an excellent target to search for low-mass companions and debris disks. Since hot white dwarfs emit mostly at ultraviolet and optical wavelengths, cooler sources of emission can be detected as excess flux in the infrared. Even a small infrared excess can be detected in the Spectral Energy Distributions (SEDs) of a white dwarf \citep{steele2011-416}.

White dwarfs with debris disks offer unique insight into the composition of the exoplanetary bodies that orbited the progenitor star. These bodies can be perturbed to pass within the tidal radius of the white dwarf, which disrupts them to form a dusty disk \citep{jura2003-584,veras2014-445}. Analysis of such systems offers the opportunity to constrain the composition of exoplanetary material \citep[e.g.][]{Klein2010}, as well as to better understand the dynamics of the last stage of evolution of planetary systems \citep{xu2014-792,wang2019-886}. Some dusty white dwarfs also exhibit a gaseous component, which often appears as double-peaked emission features \citep{gansicke2006-314,manser2020-493}.

About a dozen unresolved white dwarf brown dwarf pairs are currently known, and their evolutionary path is of great interest \citep{casewell2020-497}. If the brown dwarf was originally within $\approx$5~AU of the white dwarf progenitor, it would likely have been engulfed during its red giant phase, causing the brown dwarf to spiral inward to its current, likely tidally locked position \citep{Lagos2021}. The close orbits of such white dwarf brown dwarf pairs cause the brown dwarf to be irradiated on one hemisphere, similar to hot Jupiters. Thus, they are great analogs for studying the irradiated atmospheres of hot Jupiters orbiting main sequence stars, yet much easier to observe given the faintness of white dwarfs at infrared wavelengths \citep{zhang2017-464,lew2021-163}. These extreme atmospheric environments and evolutionary paths also present an interesting comparison with their much better understood field brown dwarf counterparts. If the brown dwarf was more widely separated from the white dwarf progenitor, they would have evolved like two single stars. There are about a dozen wide white dwarf brown dwarf pairs, which were detected via direct imaging \citep[e.g.][]{BecklinZuckerman1988,French2023}. Sometimes, the evolution path is less clear, such as the recent discovery of Gaia~0007-1605, an old hierarchical triple system with an inner white dwarf brown dwarf binary and an outer white dwarf system \citep{mansergas2022-927}.

There is much to be learned from the analysis of both white dwarf systems with debris disks, and systems with substellar companions. While many infrared excess candidate systems have been identified with photometry, differentiating between the two possible sources of excess flux can be challenging \citep{xu2020-902}. \cite{lai2021-920} and \cite{barber2014-786} outlined one method for disentangling the two possible scenarios using J band photometry. Since the dust sublimation temperatures of dust disks are around 2500~K \citep{rafikov&garmilla2013-221}, they are not expected to show significant excess at wavelengths shorter than 2 \micron. Low-mass companions in comparison can span a much larger range of temperatures, and have been shown to result in a significant J band excess. Thus, the detection of significant J band excess can rule out a dust disk in favor of a low mass companion. However, this method is limited, as a lack of J band excess does not necessarily confirm a dust disk over a companion. Near-infrared spectroscopy is a much more effective method of determining the source of excess in a white dwarf system.

In this paper, we present new infrared spectroscopic observations of seven infrared excess white dwarfs listed in Table~\ref{tab:WD-params}. They were initially identified as infrared excess candidates by cross-correlating Gaia DR2 and the unWISE catalog \citep{xu2020-902} and followed up with the \textit{Spitzer Space Telescope} \citep{lai2021-920}. Four of the systems, Gaia~J0006+2858,
WD~0145+234, Gaia J0611--6931, and Gaia~J2100+2122  were shown to have metals in their photospheres as well as circumstellar gas emissions \citep{dennihy2020-905,melis2020-905}. Two other systems, Gaia~J0052+4505 and Gaia~J0603+4518, show excess in their J band photometry. 

In Section \ref{sec:obs and red} we describe the methods used to collect the spectroscopic data with \textit{Keck/MOSFIRE}, \textit{Gemini/GNIRS}, \textit{Flamingos-2} and \textit{Keck/HIRES}, along with the data reduction steps. In Sections \ref{sec: spectral feature analysis} and \ref{sec: model fitting}, we describe our methods of analysis that include identifying spectral features and fitting the spectroscopic data. In Section \ref{discussion} we summarize the analysis of each individual system, and give the most likely source of the infrared excess for each. In Section \ref{sec:conclusion}, we conclude by outlining our results as well as the next steps forward for better understanding these and other white dwarfs with infrared excess.

\begin{deluxetable*} {l|ccccccccc}
\tabletypesize{\scriptsize}
\tablecaption{White Dwarf Parameters
\label{tab:WD-params}}
\tablecolumns{8}
\tablewidth{0pt}
\tablehead{
\colhead{Name} & 
\colhead{RA (deg)} &
\colhead{Dec (deg)} &
\colhead{Distance (pc)} &
\colhead{SpT} &
\colhead{$\mathrm{T_{eff}}$ (K)} & 
\colhead{log\textit{g} (cm s$^{-2}$)} &
\colhead{log n(Ca)/n(H)} &
\colhead{$\dot{M}_\mathrm{Ca}$(g s$^{-1}$)} &
\colhead{Ref}
}
\startdata 
Gaia~J0006+2858 & 1.644755 & 28.979655 & 151.97 $\pm$ 1.75 & DAZ & 23921 $\pm$ 335 & 8.04 $\pm$ 0.04 & -6.17 $\pm$ 0.10  & 7.9$\times$10$^6$& \citet{Rogers2022} \\
Gaia~J0052+4505 & 13.018295 & 45.092722 & 75.28 $\pm$ 0.25 & DA & 12858 $\pm$ 77 & 7.97 $\pm$ 0.01 &$<$ -8.8\tablenotemark{$\dag$}  & $<$1.6$\times$10$^4$ & \cite{kilic2020-898} \\
WD~0145+234 & 26.978382 & 23.661678 & 29.43 $\pm$ 0.02 & DAZ & 12720 $\pm$ 1000 & 8.1 $\pm$ 0.1 & -6.6 $\pm$ 0.2  & 4.3$\times$10$^6$ & \cite{melis2020-905}\\
Gaia~J0603+4518 & 90.786320 & 45.307719 & 60.30 $\pm$ 0.12 & DA & 16177 $\pm$ 323 & 8.00 $\pm$ 0.03 & $<$ -8.3\tablenotemark{$\dag$} & $<$3.8$\times$10$^4$ & \cite{gentilefusillo2021-508}\\
Gaia~J0611--6931 & 92.882367 & -69.516818 & 143.14 $\pm$ 1.05 & DAZ & 17749 $\pm$ 248  & 8.14 $\pm$ 0.04  & -6.03 $\pm$ 0.19& 8.2$\times$10$^6$ & \citet{Rogers2022}\\
Gaia~J0723+6301 & 110.823019 & 63.024055 & 137.91 $\pm$ 1.12 & DA & 18488 $\pm$ 654 & 7.92 $\pm$ 0.05 & $<$ -7.0\tablenotemark{$\dag$} & $<$7.5$\times$10$^5$ & \cite{gentilefusillo2021-508}\\
Gaia~J2100+2122 & 315.144721 & 21.382640 & 88.08 $\pm$ 0.36 & DAZ & 25565 $\pm$ 358 & 8.10 $\pm$ 0.04 & -6.22 $\pm$ 0.14& 8.0$\times$10$^6$ & \citet{Rogers2022} \\
\enddata
\tablenotetext{$\dag$}{New analysis from this paper.}
\tablecomments{Coordinates and distances are from Gaia DR3 \citep{GaiaDR3}.}

\end{deluxetable*}

\section{Observation and Data Reduction}\label{sec:obs and red}

\subsection{Near-Infrared Spectroscopy from \textit{Keck/MOSFIRE}}

\begin{deluxetable*} {l|cccccccc}
\tabletypesize{\scriptsize}
\tablecaption{Observing Log
\label{tab:observing-log}}
\tablecolumns{9}
\tablewidth{0pt}
\tablehead{
\colhead{Name} & 
\colhead{UT Date} &
\colhead{Instrument} &
\colhead{Wavelength (\micron)} &
\colhead{Exposure Time} &
\colhead{Resolution} &
\colhead{S/N} &
\colhead{Telluric} &
\colhead{Telluric $\mathrm{T_{eff}}$ (K)}
}
\startdata 
GaiaJ0006+2858 & 10-09-2020 & MOSFIRE & 1.16 --- 2.40 & J, H: 120s x 4; K: 180s x 4 & 2200 & 19 & HIP 1123 & 9579 \\
GaiaJ0052+4505 & 10-09-2020 & MOSFIRE & 1.16 --- 2.40 & J, H: 120s x 4; K: 180s x 4  & 2200 & 22 & HIP 13121 & 9620 \\
WD 0145+234    & 10-09-2020 & MOSFIRE & 1.16 --- 2.40 & J, H: 120s x 4; K: 180s x 4 & 2200 & 53 & HIP 13121 & 9620 \\
GaiaJ0603+4518 & 10-09-2020 & MOSFIRE & 1.49 --- 2.40 & H: 60s x 4; K: 180s x 4 & 2200 & 20 & HIP 13121 & 9620 \\
 & 12-12-2021 & GNIRS & 0.843 --- 2.53 & 180s x 12 & 660 & 24 & HD 26603 & 11795 \\
GaiaJ0611--6931 & 12-24-2022 & Flamingos-2 & 0.703 --- 2.65 & JH: 100s x 24, HK: 100s x 20 & 750 & 42 & HD 69784 & 9020 \\
GaiaJ0723+6301 & 02-07-2022 & GNIRS & 0.827 --- 2.53 & 300s x 16 & 540 & 7 & HD 63312 & 8475 \\
GaiaJ2100+2122 & 10-09-2020 & MOSFIRE & 1.16 --- 2.40 & J, H: 120s x 4; K: 180s x 4 & 2200 & 41 & HIP 1123 & 9579 \\
\\
GaiaJ0052+4505 & 09-22-2019 & HIRESb & 0.32 --- 0.59 &  2100s $\times$ 2& 40,000 & 12 & - & -\\
GaiaJ0603+4518 & 12-05-2019 & HIRESb & 0.32 --- 0.59 & 1200s+1100s & 40,000 & 45 & - & -\\
GaiaJ0723+6301 & 12-05-2019 & HIRESb & 0.32 -- 0.59 & 1800s $\times$ 2& 40,000 & 15 & - & -\\
\enddata

\tablecomments{Resolution is measured directly on the final spectra. S/N is estimated in regions without emission/absorption features between 1.5 and 1.7 \micron~for the infrared data and around Ca II 3933 \AA~in the optical data. Telluric standard effective temperatures from \cite{Anders2022-658}. }

\end{deluxetable*}

We used the \textit{MOSFIRE} multi-object near-infrared spectrograph \citep{McLean2010, McLean2012}, installed at the Keck\,I telescope to observe Gaia~J0006+2858,
Gaia~J0052+4505,
WD~0145+4505, 
Gaia~J0603+4518, and
Gaia~J2100+2122.  In Table~\ref{tab:observing-log} we show the details of the observations. We used \textit{MOSFIRE} in a longslit (1."0 slit width) configuration. A single photometric band is covered in each instrument setting ( $J$, $H$, or $K$). We observed all the targets, as well as the telluric standards HIP~1123 (A1 spectral type) and HIP~13121 (A0 spectral type), with a 1{\farcs}5 ABBA pattern. 

We used version 1.7.1 of PypeIt\textit{\footnote{https://github.com/pypeit/PypeIt}} to reduce all the spectra \citep{Prochaska2019, Prochaska2020}. The pipeline corrected the raw images for dark current and generated a bad-pixel mask. The edges of the slits were traced using dome flats, and a master flat was also created. PypeIt produced a wavelength calibration by use of the sky lines present in the 2D spectra. The wavelength calibration accounted for the spectral tilt across the slit. The calibrations were applied to our science frames, and the sky was subtracted using the A-B or B-A frames following \cite{Kelson2003}. The 1D science spectra were extracted from the 2D sky-corrected frames. All the white dwarfs appear as single objects in the acquisition images. We measured the Full Width at Half Maximum (FWHM) in the 2D science spectra to be 0{\farcs}7.

\begin{figure*}
    \centering
    \includegraphics[width=0.95\textwidth]{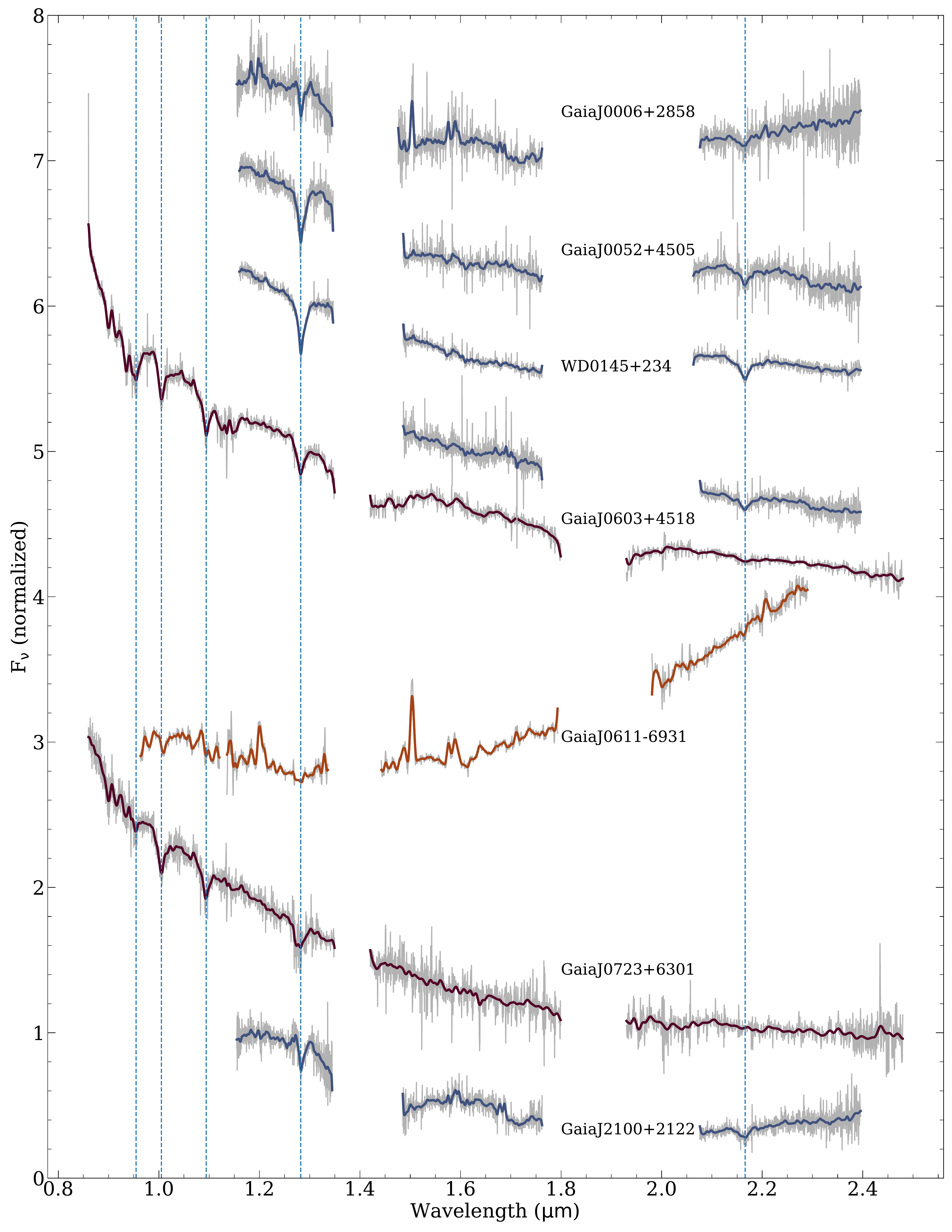}
    \caption{\textit{Flamingos-2} (orange),  \textit{MOSFIRE} (dark blue), and 
    \textit{GNIRS} (brown) spectra for each system. Spectra are divided by their median flux value and are offset by a constant from each other and are smoothed by a Gaussian filter with a standard deviation of 3 (\textit{Flamingos-2}), 10 (\textit{MOSFIRE}), and 7 (\textit{GNIRS}) pixels. White dwarf hydrogen line centroids are identified by blue dashed lines: 0.955 $\micron$ Paschen 8, 1.005 $\micron$ Paschen 7, 1.094 $\micron$ Paschen 6, 1.282 $\micron$ Paschen 5, and 2.166 $\micron$ Brackett 7. Additional emission and absorption features are detected in most white dwarfs (see Section~\ref{sec: spectral feature analysis} for further discussion). The data behind this figure are published in machine-readable format.
    \label{fig: all-spectra}
    }
\end{figure*}

Telluric correction was performed using the spectra of the corresponding telluric standard stars, as listed in Table \ref{tab:observing-log}. Hydrogen and helium features from each telluric standard's spectrum were removed individually by dividing them by a Gaussian-smoothed version of the spectrum in that region. This process removed the standard star's stellar features while preserving the telluric lines. The spectra of each telluric standard star were also divided by a blackbody curve at the star's effective temperature, leaving only telluric lines in their spectra. Telluric-corrected target spectra were flux calibrated with the near-infrared photometry reported in \cite{lai2021-920}. The final calibrated spectra for each system are shown in Figure \ref{fig: all-spectra}.

\subsection{Near-Infrared Spectroscopy from \textit{Gemini-N/GNIRS}}

Gaia~J0603+4518 and Gaia~J0723+6301 were observed with \textit{Gemini/GNIRS} \citep{elias2006-6269} via the program GN-2021B-Q-325. The short blue camera was used with the 32 l/mm grating in the cross-dispersed mode, which provides a continuous wavelength coverage of 0.8--2.5~$\mu$m. We used a 1{\farcs}0 wide slit. The standard ABBA nod pattern was adopted to facilitate sky subtraction. The exposure times are listed in Table~\ref{tab:observing-log}. Telluric standards were observed with the same configuration immediately before/after the science observations. Data reduction was performed using PypeIt \citep{Prochaska2019, Prochaska2020} and customs scripts, similar to the \textit{MOSFIRE} data reduction procedure. In the acquisition images in H band, both Gaia~J0603+4518 and Gaia~J0723+6301 appear as one single object, with a FWHM around 1{\farcs}0.

\subsection{Near-Infrared Spectroscopy from Gemini-S/Flamingos-2}

\textit{Gemini/Flamingos-2} \citep{eikenberry2004-5492} was used to observe Gaia~J0611--6931 via the program GS-2021B-Q-244. We used a 3-pixel (0{\farcs}54) wide long-slit with both the JH and HK grisms, which provides a complete wavelength coverage of 0.7--2.5~$\mu$m. The sky was clear and seeing was 0{\farcs}7. The observing log is shown in Table~\ref{tab:observing-log}. The telluric standard HD~69784 was observed immediately after the science observations. Data reduction was performed using PypeIt, similar to the \textit{MOSFIRE} and \textit{GNIRS} observations. In the acquisition image in H band, Gaia~J0611-6931 is an isloated object with a FWHM of 0{\farcs}7.

\subsection{Optical Spectroscopy from Keck/HIRES}\label{sec: optical-analysis}

\begin{figure*}
    \centering
    \includegraphics[width=\textwidth]{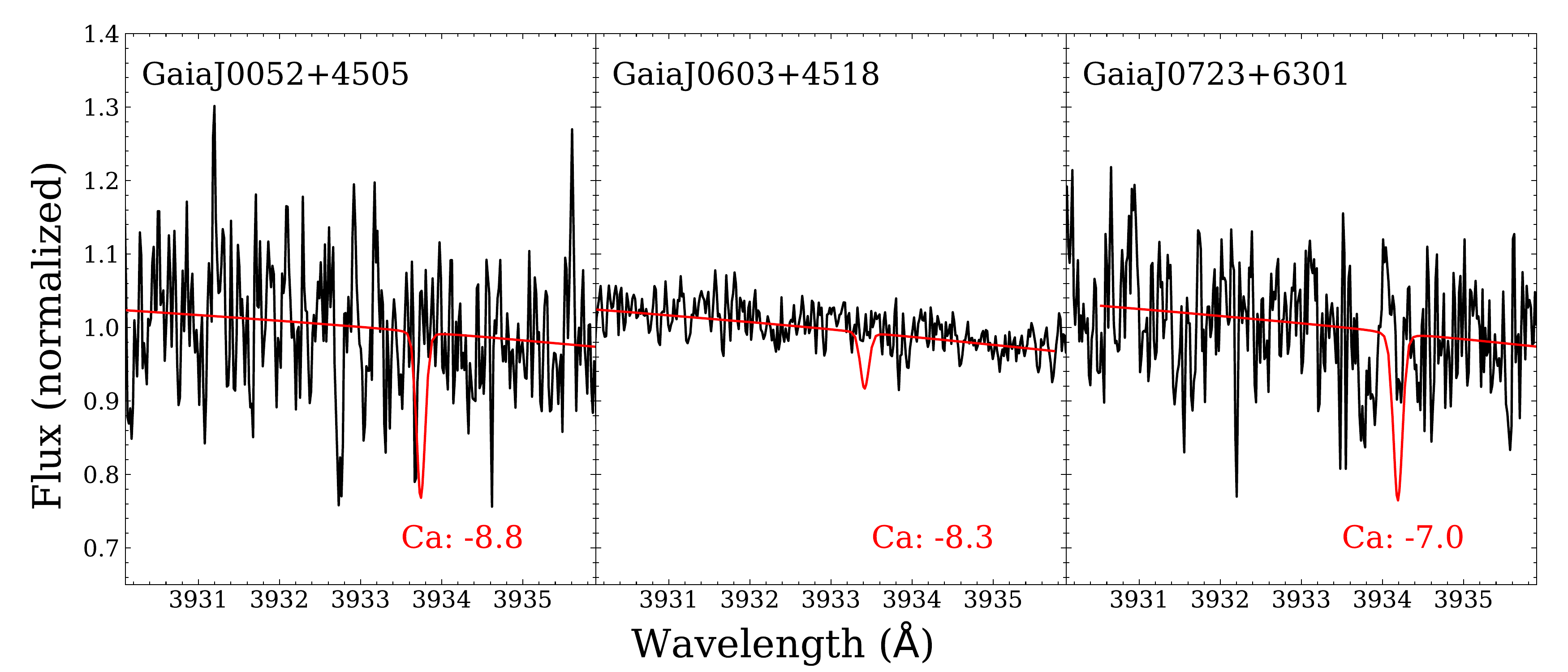}
    \caption{Keck/HIRES observations around the Ca II K line region for Gaia~J0052+4505, Gaia~J0603+4518, and Gaia~J0723+6301. The red lines are the white dwarf model that incorporates some amount of Ca, whose abundance is listed in each panel. Calcium is not detected in these systems and the abundances are upper limits. The model spectra has been shifted to match the radial velocities of each white dwarf (measured from the Balmer lines).
    \label{fig: HIRES}
    }
\end{figure*}

Gaia J0052+4505, Gaia J0603+4518, and Gaia J0723+6301 were observed in 2019 with the High Resolution Echelle Spectrometer (HIRES) on the Keck I \citep{vogt1994-2198} via program 2019B$\_$N072, as part of our effort to characterize Gaia white dwarfs with an infrared excess. The observing log is in Table~\ref{tab:observing-log} and clouds were variable during the night. Data reduction was performed using the MAKEE package and then continuum normalized with IRAF, following our previous HIRES observations \citep{xu2016-816}. In all three stars, we identified broad Balmer lines from the white dwarf, and their radial velocities are consistent to within $\approx$ 30~km~s$^{-1}$ in the consecutive spectra. We did not find any narrow absorption features from heavy elements or narrow emissions from potential companions. We focused on the Ca II-K line around 3933~{\AA} and following procedures described in \citet{Rogers2022}. An upper limit in the equivalent width (EW) of Ca II K line was derived, which is 32~m{\AA}, 10~m{\AA}, and 28~m{\AA} for Gaia~J0052+4518, Gaia~J0603+4505, and Gaia~J0723+6301, respectively. We then computed white dwarf model atmospheres \citep{dufour2007-663} and derived an upper limit on the calcium abundance and mass accretion rate, as shown in Figure \ref{fig: HIRES} and Table \ref{tab:WD-params}. 

The remaining four white dwarfs, i.e. Gaia~J0006+2858,
WD~0145+234, Gaia~J0611--6931,  and Gaia~J2100+2122, are heavily polluted and the abundance analysis have been reported in \citet{Rogers2022} (see Table~\ref{tab:WD-params}).

\section{Spectral Feature Analysis}\label{sec: spectral feature analysis}

In hot DA (hydrogen-dominated) white dwarfs, like those in this sample, there are only a few hydrogen lines from the star's atmosphere in the spectra, making features from a disk or companion easily detectable. The infrared spectra of debris disks and brown dwarf companions have different characteristics, which can be used to assess the source of infrared excess. Previous near-infrared observations of white dwarfs with debris disks show that these systems have a featureless continuum from heated dust \citep{kilic2008-136,melis2011-732}. Mid-infrared studies using the \textit{Spitzer/IRS} have detected broad silicate emissions centered at 10 $\mathrm{\mu m}$, which have been modeled to originate from an optically thin region of a dust disk \citep{reach2005-635,jura2007-133}. On the other hand, for white dwarf brown dwarf pairs we expect to see broad absorption bands from molecules in the brown dwarf's atmosphere \citep{mclean2003-596,casewell2020-497,lew2021-163}. In the infrared, prominent absorption features come from $\mathrm{H_2O}$ in the J band, and CO in the K band \citep{mclean2003-596}.

In all the systems, with the exception of Gaia~J0611--6931, we have detected broad hydrogen absorption lines from the photosphere of each white dwarf, as shown in Figure~\ref{fig: all-spectra}. These lines were consistent with our white dwarf models \citep{dufour2007-663} calculated from the $\mathrm{T_{eff}}$ and log\textit{g} for each system. Interestingly, we did not detect Paschen or Brackett lines in Gaia~J0611--6931, even though the Balmer series are clearly detected and well modeled in the optical data \citep{Rogers2022}. Gaia~J0611--6931 shows strong emission features (see Section~\ref{sec:J0611}) throughout the near-infrared, and we suspect there are additional emission features around the Paschen and Brackett lines, complicating their detections.

\subsection{Gaseous Debris Emissions}

\begin{figure*}
    \centering
    \includegraphics[width=0.48\textwidth]{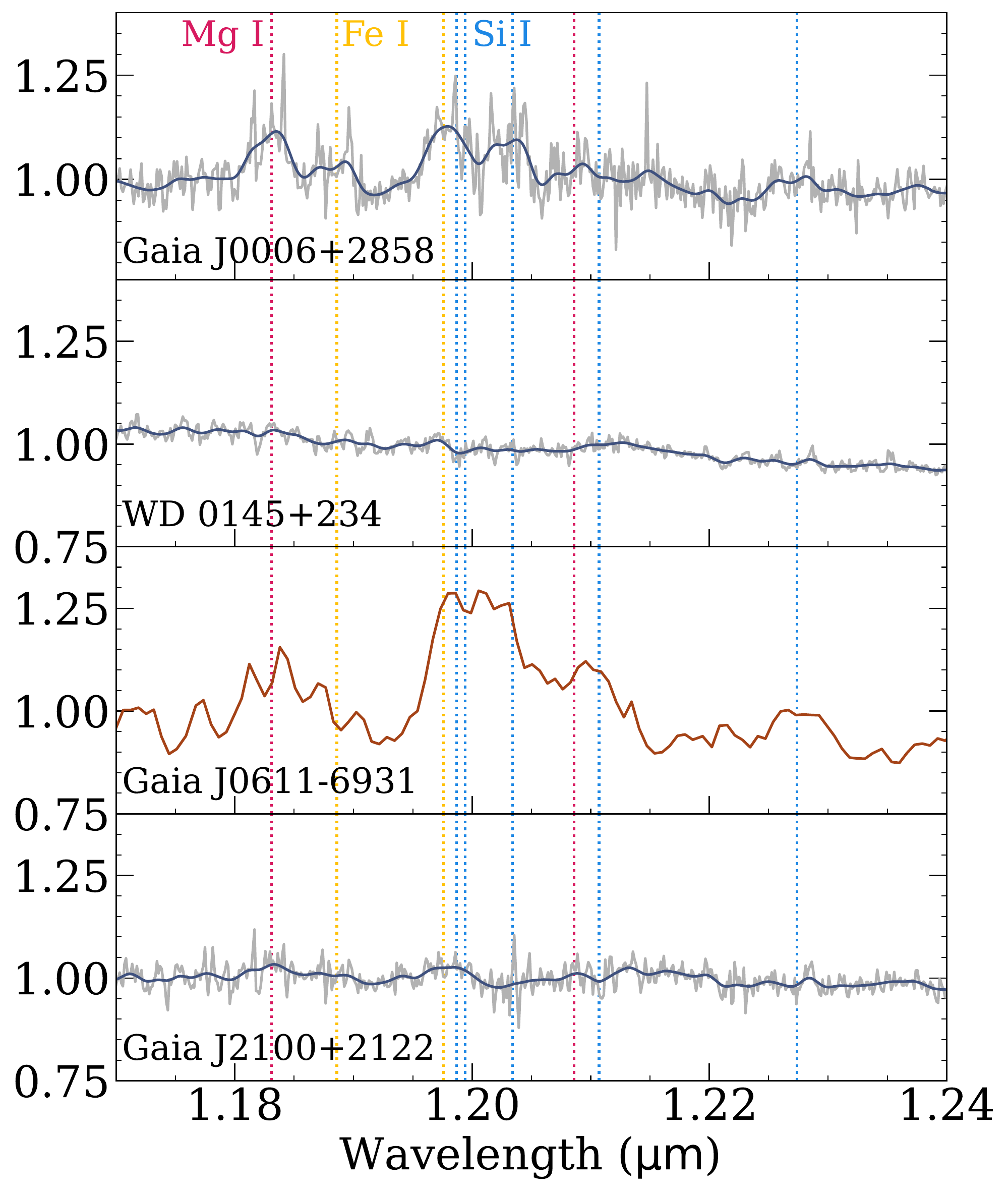}
    \includegraphics[width=0.48\textwidth]{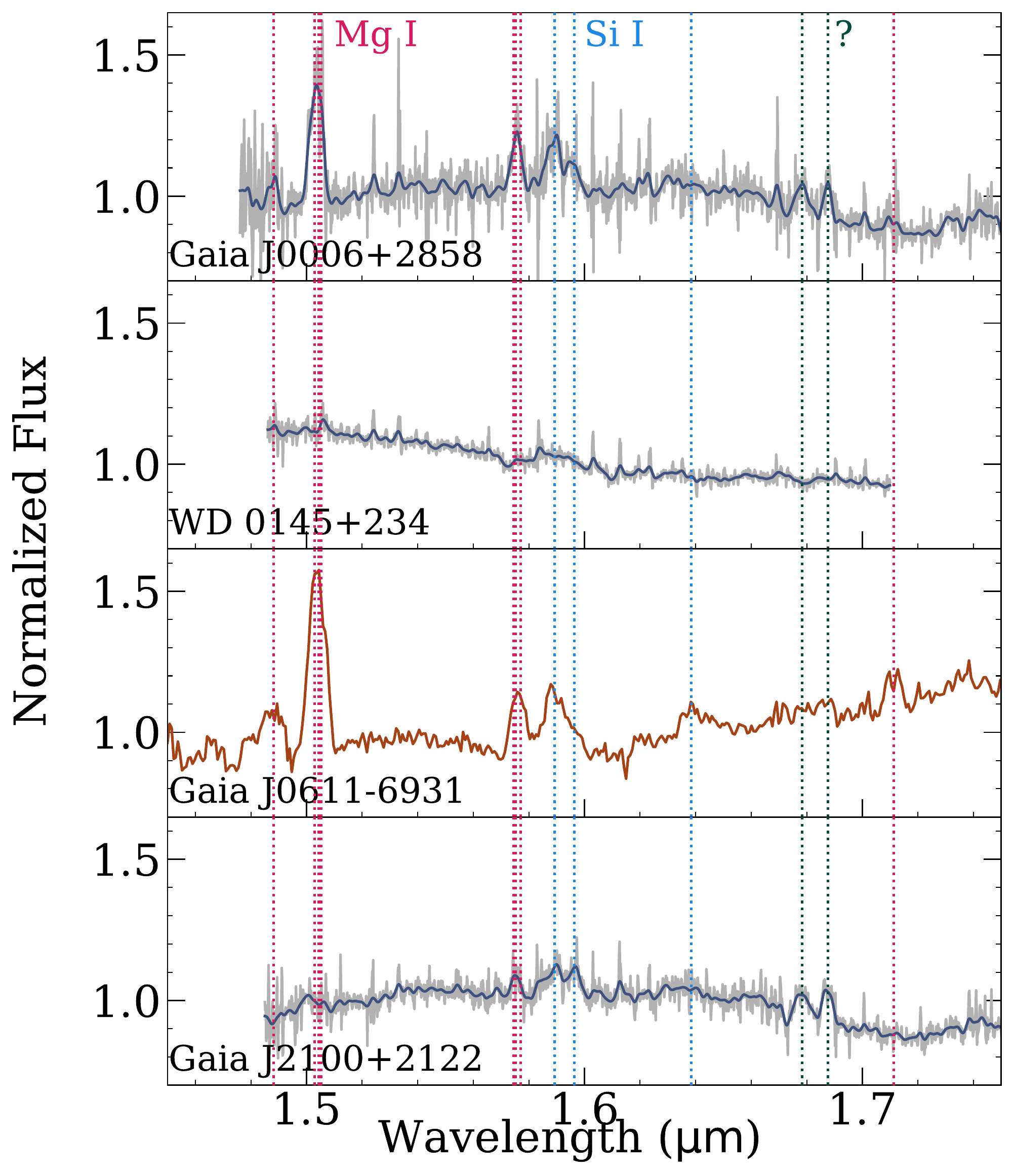}
    \caption{Spectra of near-infrared emission lines for each of the disk systems. Gray spectra shows the observed \textit{MOSFIRE} data, while dark blue spectra show that data smoothed by a Gaussian filter with a standard deviation of 5. The Gaia~J0611--6931 \textit{Flamingos-2} spectra are shown in orange. Metal emissions from Mg I, Fe I, and Si I are detected in Gaia~J0611--6931, Gaia~J0006+2858 and Gaia~J2100+2122. We did not identify metal emissions in the \textit{MOSFIRE} spectrum of WD 0145+234. The vertical dotted lines mark the centeroid of each emission feature, as listed in Table~\ref{tab:emission lines}. Magenta lines mark Mg I emissions, yellow lines mark Fe I emissions, blue mark Si I emissions, and dark green mark unknown emissions.
    \label{fig: emission_figure}
    }
\end{figure*} 

For the first time, we have detected strong infrared emission features in Gaia~J0006+2858,
Gaia~J0611--6931, and
Gaia~J2100+2122, as shown in Figure \ref{fig: emission_figure}. These three systems also host a range of metal emissions in their optical spectra, such as Ca II, O I, Fe II, Mg I, Si I, and Na I \citep{melis2020-905,dennihy2020-905}. We used the optical line identifications and transition properties to help with our own line identification of the infrared emission lines. We used atomic line lists from \cite{vanhoof2018-6}\footnote{\url{https://www.pa.uky.edu/~peter/newpage/}} to query transition properties (transition energy levels, Einstein coefficients, and oscillator strengths). Comparing transition characteristics from lines identified in previous studies, we tried to find the most-likely candidate for each observed feature. As an additional confirmation, we checked if there were other lines of the same element that we would also expect to see given the properties of the candidate transition. If there were other expected transitions within the wavelength range of our data that we did not observe, then we rejected that candidate. Due to the lower resolution of the infrared spectra, many of the emission features appear as one broad single peak, as opposed to double peaked features in the optical observations \citep{melis2020-905,dennihy2020-905}. We fit each observed line with a Gaussian function to determine the feature's centroid. We calculated line properties including the full width at zero intensity (FWZI), equivalent width (EW), and radial velocity (RV). Table \ref{tab:emission lines} reports the properties of the emission features that we identified for each system.

In Gaia~J0006+2858,
Gaia~J0611--6931, and
Gaia~J2100+2122, we identified emission features consistent with the species observed in previous optical studies including Mg I, Fe I, and Si I. These emission lines are highlighted for each system in Figure \ref{fig: emission_figure}. The FWZI values are mostly consistent within each system, with the exception of a few lines that are likely a blend of several transitions. For example, the Mg I line at 1.5029 \micron~had a much larger FWZI than any of the other lines observed in Gaia~J0006+2858. We attribute this to other nearby Mg I lines (1.5044 \micron~ \& 1.5051 \micron) that could be blended with the line. Also, the two sets of close Si I lines (1.1987 \& 1.2034 \micron, 1.5892 \& 1.5964 \micron) in Gaia~J0611--6931 are blended, so we report combined measurements of FWZI and EW for them. The 1.1976 \micron~ Fe I line is blended with the 1.1987 \micron~ and 1.2034 \micron~ Si I lines in the Gaia~J0006+2858, Gaia~J0611--6931, and Gaia~J2100+2122 spectra, so we list that Fe I line as blended in Table \ref{tab:emission lines}. There are also two unidentified emission lines that we list in Table \ref{tab:emission lines}.

The RV values that we measured among each system varied between different transitions, and had higher uncertainties than those reported in optical studies of these systems \citep{melis2020-905,dennihy2020-905}. We attribute this to the low spectral resolutions in our data compared to optical observations of these systems, as well as the blended nature of many of the lines we observed, causing the centroids of individual lines and velocity shifts to be blurred. For these reasons, we did not compare our measured RV values with those reported in \cite{melis2020-905} and \cite{dennihy2020-905} as a validity check for our line identifications.

\begin{deluxetable*} {ccccccccccccc}
\tabletypesize{\scriptsize}
\tablecaption{Observed Emission Lines
\label{tab:emission lines}}
\tablecolumns{15}
\tablewidth{0pt}
\tablehead{
\colhead{Transition} & 
\colhead{$\mathrm{E_{high}}$} & \colhead{log(\textit{gf})} &
\multicolumn{3}{c}{Gaia~J0006+2858} &
\multicolumn{3}{c}{Gaia~J0611--6931} &
 \multicolumn{3}{c}{Gaia~J2100+2122} \\
& & & \colhead{EW} & \colhead{FWZI} & \colhead{RV} & \colhead{EW} & \colhead{FWZI} & \colhead{RV} & \colhead{EW} & \colhead{FWZI} & \colhead{RV} \\ (Vacuum $\mathrm{\micron}$) & \colhead{({eV})} & & \colhead{({\AA})} & \colhead{(km s$^{-1}$)} & \colhead{(km s$^{-1}$)} & \colhead{({\AA})} & \colhead{(km s$^{-1}$)} & \colhead{(km s$^{-1}$)} & \colhead{({\AA})} & \colhead{(km s$^{-1}$)} & \colhead{(km s$^{-1}$)}
}
\startdata 
\textbf{Magnesium} &  &  & & & \\
Mg I 1.1831 & 5.3937 & -0.3319 & $4.2\pm0.8$ & $1551\pm220$ & $-10\pm46$ & $10\pm2$ & $3226\pm560$ & $228\pm109$  &  $1.2\pm0.5$ & $1686\pm503$ & $20\pm101$\\
Mg I 1.2086 & 6.7790 & 0.3733 & - & - & -
& $8.6\pm1.9$ & $2536\pm464$ & $248\pm109$  & $0.3\pm0.1$ & $724\pm115$ & $22\pm25$ \\
Mg I 1.4882\tablenotemark{$\dag$} & 6.7790 & 0.6978 & $4.9\pm2.1$ & $1452\pm494$ & b & $15\pm2$ & $3558\pm447$ & b  & - & - & - \\
Mg I 1.5029 & 5.9328 & 0.3577 & $24\pm2$ & $1973\pm125$ & b & $40\pm3$ & $2400\pm122$ & b & - & - & - \\
Mg I 1.5044 & 5.9320 & 0.1355 & b & b & b & b & b & b & - & - & - \\
Mg I 1.5052 & 5.9320 & -0.3422 & b & b & b & b & b & b & - & - & - \\
Mg I 1.5745 & 6.7190 & -0.2118 & $6.9\pm1.3$ & $1198\pm179$ & b & $12\pm1$ & $1918\pm139$ & b & $2.1\pm0.6$ & $1019\pm233$ & b \\
Mg I 1.5753 & 6.7190 & 0.1402 & b & b & b & b & b & b & b & b & b \\
Mg I 1.5770 & 6.7190 & 0.4108 & b & b & b & b & b & b & b & b & b \\
Mg I 1.7113 & 6.1182 & 0.0648 & $2.4\pm1.1$ & $1491\pm549$ &  $-201\pm107$ & $5.8\pm1.4$ & $1928\pm363$ & $-35\pm86$  & - & - & - \\
\textbf{Silicon} &  &  & & &  & & & \\
Si I 1.1987 & 5.9639 & 0.1775 & $4.6\pm0.8$ & $1395\pm257$ & b & $26\pm3$ & $3592\pm277$ & b & $1.5\pm0.4$ & $1702\pm390$ & b \\
Si I 1.1994 & 5.9537 & -0.1720 & b & b & b & b & b & b & b & b & b \\
Si I 1.2034 & 5.9840 & 0.4181 & $3.2\pm0.8$ & $1456\pm291$ & b & b & b & b & - & - & - \\
Si I 1.2107 & 5.9537 & -0.3948 & b & b & b & b & b & b & - & - & - \\
Si I 1.2274 & 5.9639 & -0.4379& $1.8\pm0.7$ & $1818\pm520$ & $-1.4\pm103$ & $4.9\pm0.8$ & $2165\pm266$ & $73\pm51$  & - & - & - \\
Si I 1.5893 & 5.8624 & -0.0068 & $6.6\pm1.8$ & $1561\pm327$ & $-26\pm66$ & $24\pm2$ & $4011\pm236$ & $94\pm47$ & $5.8\pm1.1$ & $2501\pm384$ & $25\pm77$ \\
Si I 1.5964 & 6.7606 & 0.1976  & $3.1\pm1.6$ & $1309\pm524$ & $-131\pm100$ & b & b & b & $3.2\pm0.7$ & $1204\pm187$ & $65\pm39$ \\
Si I 1.6385 & 6.6192 & -0.2718  & - & - & - & $9.2\pm1.2$ & $3697\pm396$ & $165\pm75$ & - & - & - \\
Si I 1.6386 & 6.7206 & -0.4225  & - & - & - & b & b & b & - & - & \- \\
\textbf{Iron}\tablenotemark{$\ddag$} &  &  & & &  & & & \\
Fe I 1.1886 & 3.2410 & -1.6676  & $2.6\pm0.8$ & $1211\pm223$ & $-100\pm48$ & b & b & b & $0.7\pm0.2$ & $933\pm242$ & $-12\pm105$ \\
Fe I 1.1976 & 3.2112 & -1.4828  & b & b & b & b & b & b & b & b & b \\
\hline
\textbf{Unidentified} &  &  & & &  & & &\\
 & & & EW & FWZI & $\lambda$& EW & FWZI & $\lambda$ & EW & FWZI & $\lambda$\\
  & & & ($\mathrm{\AA}$) & (km s$^{-1}$) &  ($\mathrm{\micron}$) &($\mathrm{\AA}$) & (km s$^{-1}$) &  ($\mathrm{\micron}$) &($\mathrm{\AA}$) & (km s$^{-1}$) &  ($\mathrm{\micron}$) \\
? &  &  & $4.4\pm1.2$ & $1361\pm320$ & $1.67834 \pm 0.0032$ & - & - & -  & $5.0\pm0.9$ & $1737\pm237$ & $1.67842 \pm 0.0027$ \\
? &  &  & $4.2\pm1.0$ & $825\pm149$ & $1.68767 \pm 0.0012$ & - & - & -  & $5.9\pm0.8$ & $1371\pm139$ & $1.68750 \pm 0.0013$ \\
\enddata
\tablenotetext{$\dag$}{The measurements for this feature are a blend of up to five Mg I emissions with the same $\mathrm{E_{high}}$ located between 1.4881 \micron~ and 1.4882 \micron. We list the log(\textit{gf}) of the highest probability transition.}
\tablenotetext{$\ddag$}{The Fe I 1.1886 line we identified is blended with the Mg I 1.1831 line in Gaia~J0611--6931. The Fe I 1.1976 line we identified is blended with the Si I 1.1987 and 1.1994 lines in all three systems.}
\tablecomments{We did not identify any emission features from GaiaJ0052+4505, WD 0145+234, GaiaJ0723+6301, or GaiaJ0603+4518, they have been omitted from this table. We could not disentangle some sets of nearby emissions, so we denote each line that is blended with the line listed above it in that wavelength region as "b". Any RV measurement of a blended line is also listed as "b".
}

\end{deluxetable*}

\subsection{Brown Dwarf Features}

In the spectra of Gaia~J0052+4505 and Gaia~J0603+4518, we identified broad absorption bands consistent with molecular features from the atmospheres of brown dwarfs. In order to assess these broad spectral features, the flux contribution from the white dwarf had to first be removed. White dwarf model spectra were calculated with parameters listed in Table~\ref{tab:WD-params} and calibrated to flux units using optical photometry from \textit{PanSTARRS} \citep{chambers2019-panstarrs} or \textit{SkyMapper} \citep{onken2019-skymapper}. To subtract the white dwarf model from the observed spectroscopy, the model spectrum was interpolated to the observed wavelength grid of each spectrum.

\begin{figure*}
    \centering
    \includegraphics[width=0.99\textwidth]{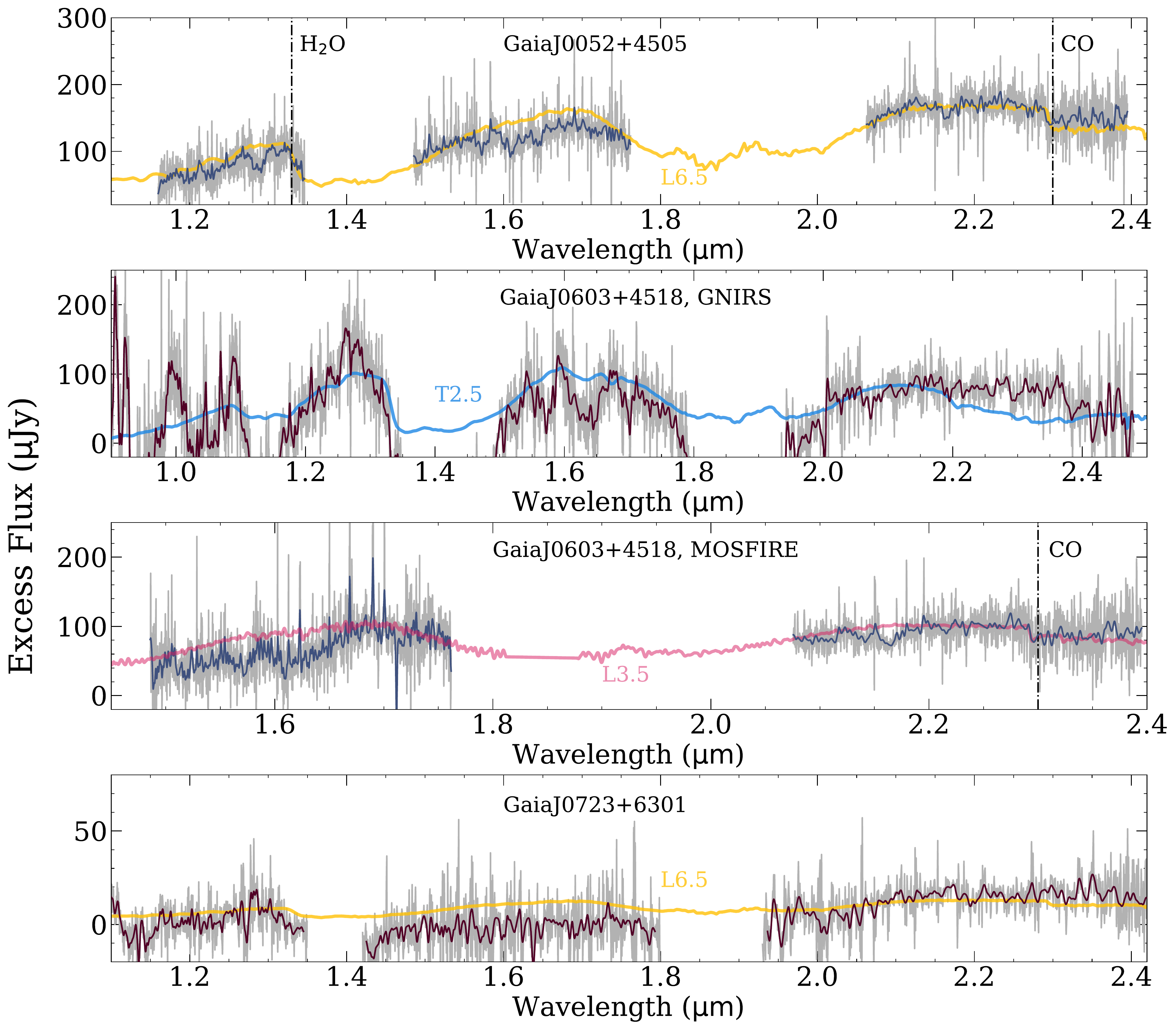}
    \caption{Near infrared spectra for the two putative brown dwarf companion systems (Gaia~J0603+4518, Gaia~J0052+4505) and Gaia~J0723+6301. The white dwarf contribution has been subtracted, isolating the excess flux. The original spectrum is shown in grey, along with a Gaussian-smoothed version in color. Each spectrum is shown with a best-fitting brown dwarf template spectrum for comparison, as described in Section \ref{sec: BD template comparison}. Vertical black dotted lines indicate brown dwarf atmospheric absorption features. 
    \label{fig: BD comparison}
    }
\end{figure*}

In Figure \ref{fig: BD comparison} we identify an $\mathrm{H_2O}$ absorption band in the \textit{MOSFIRE} spectra of Gaia~J0052+4505. This feature is observed longward of 1.33 $\mathrm{\mu m}$, and is present in late M dwarfs and later dwarf types \citep{mclean2003-596}. In the K band, we clearly detected the CO band head in the \textit{MOSFIRE} spectra of both Gaia J0052+4505 and  Gaia J0603+4518. This feature is known for its strong absorption longward of 2.3 $\mathrm{\mu m}$, and is prominent in the spectra of M dwarfs, L dwarfs, and early T dwarfs \citep{mclean2003-596}. However, we did not observe this feature in the  \textit{GNIRS} spectrum of the same object. We defer the discussion about this to Section~\ref{sec: J0603}.

The \textit{GNIRS} spectrum of Gaia J0723+6301 was mostly consistent with the white dwarf model. In fact, the infrared excess does not start until the K band, as shown in Figure~\ref{fig: BD comparison}.

\section{Model Fitting} \label{sec: model fitting}

\subsection{Dust Disk Model Fitting \label{sec: disk model fitting}}

The infrared excesses were fit using the dust disk model outlined in \cite{jura2003-584}. This model describes a flat, opaque ring of blackbody-emitting dust, where the dust's effective temperature depends only on its orbital radius. Such a dust ring is characterized by three disk parameters: inner radius ($R_{in}$), outer radius ($R_{out}$), and inclination angle ($i$). Given the degeneracies among these three parameters in previous applications \citep[e.g.][]{jura2007-133}, we decided to use a Markov chain Monte Carlo (MCMC) technique to fully explore the parameter space, using the \textit{emcee} Python module \cite{foreman-mackey2013-125}. The two parameters describing disk radii ($R_{in}$ and $R_{out}$) were uniformly sampled in logarithmic space, to avoid potential sampling bias from the non-linear relationship between the disk radius and temperature. Inclination was uniformly sampled between 0-90\textdegree. A lower boundary was placed on $R_{in}$ at a sublimation radius where the dust ring temperature is 2500~K, a temperature above which all dust would be expected to sublimate \citep{steckloff2021-913}. An upper boundary was placed on $R_{out}$ at a tidal radius of 300$\mathrm{R_{WD}}$, a loose constraint for the Roche Limit expected for these white dwarfs \citep{steckloff2021-913}. The MCMC chains were run using 100 walkers, each with 10,000 steps including a 5,000 step burn-in phase.

We fit each system using the new infrared spectroscopic data along with \textit{Spitzer} 3.6 and 4.5~$\mathrm{\mu m}$ photometry. Given the large difference in wavelength coverage and sampling between the photometry and spectroscopy, in order to combine the two for fits, a weighting factor was added to the likelihood function function of the MCMC framework. Each spectroscopy and photometry point was given a weight proportional to its wavelength coverage. We report the median values for each model parameter from these fits with 1$\sigma$ uncertainties in Table \ref{tab:MCMC_table}. The best fit models for the dust disks are shown in Figure~\ref{fig: dust_fits}. We also show the corner plots for these fits in Figures \ref{fig: disk-corners} and \ref{fig: BD-corners}.

We also performed the same MCMC fitting routine using the near-infrared (NIR) photometry in Table \ref{tab:photometry} to compare with the results of our spectroscopic fits. We found that there was very little difference between the photometry-only fits (JHK and \textit{Spitzer} photometry) and the fits with NIR spectroscopy and \textit{Spitzer} photometry. We believe this is due to the similar weighting of NIR photometry points and the spectral data, which when fit with the \textit{Spitzer} data, resulted in a similar ability to constrain the model parameters. We tested fits with only NIR photometry or spectroscopy, excluding \textit{Spitzer} points, and found in this scenario that spectroscopy was able to constrain the model parameters better than photometry. Given this, we believe the best combination of data to fit is the NIR spectroscopy with the \textit{Spitzer} data.

In our analysis of the MCMC dust disk fits, we noted several patterns among the model parameters ($i$, $R_{in}$, and $R_{out}$) across each of the systems we fit. The inner radius was always the best constrained parameter in our fits of all the systems. This is expected, as the near-infrared is most sensitive to the warmest region of the dust disk near the inner radius. We found that for all of the disk systems, the temperatures of the inner disk radius were within a reasonable range for dust sublimation, falling between 1200 K and 1900 K.

The spread in the posterior distributions of disk inclinations was usually large; the only exception is Gaia~J0611--6931 whose large excess flux forced a face-on disk. The spread is due to the inclination's degenerate nature with the inner and outer radii of the disk. In most of the systems, again with the exception of Gaia~J0611--6931, we saw a correlation between $i$ and $R_{out}$, where larger inclination angles corresponded to larger values of $R_{out}$. We saw the opposite correlation between $i$ and $R_{in}$, where larger inclination angles corresponded to smaller values of $R_{in}$. These correlations are illustrated in the corner plots in Figure \ref{fig: disk-corners}. Conceptually, the correlation in these parameters is to be expected, as a more face-on disk would require a smaller emitting area to model the same SED as a more edge-on disk with a larger emitting area.

The posterior distributions for the outer radius parameter consistently showed a peak at a smaller value, with an unconstrained tail going to larger radii. This was expected due to the lack of longer wavelength data. The outer radius is the coolest part of the disk, which the mid-infrared is much more sensitive to than the near-infrared. If there were available spectral data for these systems at wavelengths longer than our \textit{Spitzer} photometry, our ability to constrain this parameter would greatly improve.

When we compared the posterior distributions for each disk parameter between the disk systems and the likely brown dwarf companion systems, we found some differences with the best fitting disk radii that can be seen in the corner plots in Figures \ref{fig: disk-corners} and \ref{fig: BD-corners}. The $R_{in}$ distribution for each of the likely brown dwarf companion systems was close to the dust sublimation limit. This can be seen in Figure \ref{fig: BD-corners} with the sharp cutoff on the lower end of each $R_{in}$ distribution. Hot disks were required to fit the beginning of excess flux at shorter wavelengths, a trait of brown dwarf companion systems \citep{lai2021-920}. We also found that these systems were fit by narrower disks, with the $R_{out}$ distributions close to the best fitting $R_{in}$.

\begin{deluxetable*} {lcccccccccc}
\tabletypesize{\scriptsize}
\tablecaption{Median Dust and Gas Disk Model Parameters and Best Fitting Brown Dwarf Companion
\label{tab:MCMC_table}}
\tablecolumns{10}
\tablewidth{0pt}
\tablehead{
\colhead{Name} & 
\colhead{Inclination} &   
\colhead{$\mathrm{R_{dust,in}}$} &
\colhead{$\mathrm{R_{dust,out}}$} &
\colhead{$\mathrm{v_{in}sin}i$} & 
\colhead{$\mathrm{v_{out}sin}i$} &
\colhead{$\mathrm{R_{gas,in}}$} &
\colhead{$\mathrm{R_{gas,out}}$} &
\colhead{BD Fit} &
\colhead{Likely IR Source} \\ & \colhead{(deg)} & \colhead{($\mathrm{R_{WD}}$)} & \colhead{($\mathrm{R_{WD}}$)} & \colhead{(km s$^{-1}$)} & \colhead{(km s$^{-1}$)} & \colhead{($\mathrm{R_{WD}}$)} & \colhead{($\mathrm{R_{WD}}$)}
}
\startdata 
Gaia~J0006+2858 & $66^{+6}_{-33}$ & $20^{+4}_{-4}$ & $64^{+105}_{-24}$ & 545 & 230 & $22^{+8}_{-13}$ & $92^{+32}_{-55}$ & L6.5 $\pm$ 4.5 & Disk \\
Gaia~J0052+4505 & $42^{+23}_{-28}$ & $5.7^{+0.7}_{-0.5}$ & $7.4^{+1.2}_{-0.8}$\ & - & - & - & - & L6.5 $\pm$ 4.0 & BD \\
WD~0145+234 & $30^{+21}_{-20}$ & $13^{+1}_{-1}$ & $24^{+4}_{-2}$ & 640 & 358 & $8^{+8}_{-7}$ & $22^{+23}_{-20}$ & L6 $\pm$ 4 & Disk \\
Gaia~J0603+4518 &  &  &  &  &  \\
\textit{MOSFIRE} & $73^{+12}_{-43}$ & $11^{+2}_{-2}$ & $15^{+6}_{-3}$ & - & - & - & - & L3.5 $\pm$ 3.5 & BD \\
\textit{GNIRS} & $73^{+13}_{-40}$ & $11^{+4}_{-3}$ & $15^{+8}_{-4}$ & - & - & - & - & T2.5 $\pm$ 2.5 & BD \\
Gaia~J0611--6931 & $3.6^{+4.0}_{-2.6}$ & $10.0^{+0.2}_{-0.2}$ & $165^{+83}_{-57}$ & 650 & 313 & $0.28^{+0.20}_{-0.27}$ & $1.1^{+0.8}_{-1.0}$ & L7 $\pm$ 4 & Disk \\
Gaia~J0723+6301 & $75^{+10}_{-42}$ & $23^{+8}_{-8}$ & $39^{+74}_{-12}$ & - & - & - & - & L6.5 $\pm$ 4 & ? \\
Gaia~J2100+2122 & $60^{+15}_{-35}$ & $33^{+4}_{-6}$ & $57^{+42}_{-9}$ & 370 & 230 & $48^{+26}_{-36}$ & $117^{+64}_{-88}$ & L6 $\pm$ 4 & Disk\\
\enddata

\tablecomments{Dust disk model fits included NIR spectra with \textit{Spitzer} photometry. Brown dwarf companion fits included NIR spectra only. Errors for the disk model parameters are the 84.1 and 15.9 percentile values (1$\sigma$) from each parameter's MCMC posterior distribution. Gas disk calculations are omitted for systems lacking metal emissions. Details for each model are described in Section~\ref{sec: model fitting}. Gaia J0603+4518 was observed with \textit{MOSFIRE} and \textit{GNIRS}, which are reported separately in this table.}

\end{deluxetable*}

\begin{deluxetable*} {l|ccccccc}
\tabletypesize{\scriptsize}
\tablecaption{Infrared photometry
\label{tab:photometry}}
\tablecolumns{6}
\tablewidth{0pt}
\tablehead{
\colhead{Name} & 
\colhead{J (mag)} &
\colhead{H (mag)} &
\colhead{K (mag)} &
\colhead{IRAC Ch1 (mag)} &
\colhead{IRAC Ch2 (mag)} &
}
\startdata 
Gaia~J0006+2858 & 16.80 $\pm$ 0.06 & 16.76 $\pm$ 0.05 & 16.13 $\pm$ 0.08 & 14.96 $\pm$ 0.06 & 14.38 $\pm$ 0.06 \\
Gaia~J0052+4505 & 16.02 $\pm$ 0.05 & 16.03 $\pm$ 0.03 & 15.61 $\pm$ 0.04 & 15.43 $\pm$ 0.06 & 15.39 $\pm$ 0.06 \\
WD~0145+234 & 14.32 $\pm$ 0.11 & 14.26 $\pm$ 0.08 & 13.79 $\pm$ 0.08 & 12.83 $\pm$ 0.06 & 12.43 $\pm$ 0.06 \\
Gaia~J0603+4518 & 15.40 $\pm$ 0.04 & 15.43 $\pm$ 0.02 & 15.33 $\pm$ 0.03 & 14.97 $\pm$ 0.06 & 14.92 $\pm$ 0.06 \\
Gaia~J0611--6931 & 17.23 $\pm$ 0.06 & 16.69 $\pm$ 0.05 & 15.44 $\pm$ 0.05 & 14.13 $\pm$ 0.06\tablenotemark{$\dag$} & 13.34$\pm$ 0.06\tablenotemark{$\dag$} \\
Gaia~J0723+6301 & 17.36 $\pm$ 0.43 & 17.37 $\pm$ 0.35 & 17.04 $\pm$ 0.15 & 16.39 $\pm$ 0.06 & 16.05 $\pm$ 0.06 \\
Gaia~J2100+2122 & 15.75 $\pm$ 0.04 & 15.71 $\pm$ 0.03 & 15.34 $\pm$ 0.05 & 14.04 $\pm$ 0.06 & 13.54 $\pm$ 0.06 \\
\enddata

\tablenotetext{$\dag$}{\textit{WISE} magnitudes, as the \textit{Spitzer} data was marked as potentially spurious in \cite{dennihy2020-905}.}
\tablecomments{Photometry is from \cite{lai2021-920}, except for Gaia J0611--6931, which is from \cite{dennihy2020-905}. The JHK magnitudes are reported in the MKO system.}

\end{deluxetable*}

\subsection{Gas Disk Model Fitting \label{sec: gas model fitting}}

Using metal emissions lines observed in four of our systems (Gaia~J0006+2858, WD~0145+234, Gaia~J0611--6931, Gaia~J2100+2122), we calculated the inner and outer radius of the emitting gas region for each system. In order to make this calculation, we assumed that the gas in each system is orbiting the white dwarf in a circular disk and has the same inclination in our line of sight as the dust. Disk formation models suggest that such assumptions are reasonable, as gaseous material can be formed either from impacts between bodies within the dust disk or from dust sublimation at the hot inner edge of the disk, spreading throughout the disk from there \citep{jura2008-135,melis2010-722,rafikov2011-416,rafikov2011-732,bochkarev2011-741,hartmann2011-530,hartmann2016-593,metzger2012-423,bear2013-19,kenyon2017-850}.

With our assumptions about the geometry of the emitting gas disks in our systems, we related observed properties of the emissions to the inner and outer radius of each gas disk. The fastest-moving gas located at the disk's inner radius is related to half of the full width of a gas emission line, which we list as $\mathrm{v_{in}sin}i$ in Table \ref{tab:MCMC_table}. On the other hand, when double-peaked line profiles are observable, peak separation is about twice the velocity at the outer radius of emitting gas ($\mathrm{v_{out}sin}i$).

While \cite{melis2020-905} measured these properties for the four gas-emitting systems in this paper, they warned against drawing such conclusions about the disk geometry without sufficiently modeling the dust portion and viewing angle of the debris disk. Our MCMC fitting technique of the infrared spectra described in Section \ref{sec: disk model fitting} does just this. The resulting posterior distributions for disk inclinations illustrated in Figure \ref{fig: disk-corners} provide us with a reasonable uncertainty range of inclinations based on the observed infrared excess.

While we measured the full widths of the gas emissions we observed in our infrared spectra, which are listed in Table \ref{tab:emission lines}, we decided to use the reported values in \cite{melis2020-905} for our gas disk calculation because the optical emission lines are well resolved. We listed these values in Table \ref{tab:MCMC_table}.

We found that for three of our four gas emissions systems (Gaia~J0006+2858, WD~0145+234, and Gaia~J2100+2122), the gas inner and outer radii were all consistent with our modeled dust disk properties reported in Table \ref{tab:MCMC_table}. The gaseous material appears to spatially coincide with the dust disk, as has been found in other gaseous systems \citep{melis2010-722}.

The calculated gas inner and outer radii of one system, Gaia~J0611--6931, stand out as being unreasonably small and in disagreement with the best-fitting dust disk properties for the system. This can be attributed to our dust disk model's inability to sufficiently reproduce the strong brightness of infrared excess observed in this system. Our model was forced to a face-on disk for this system, meaning the inclination range we used for our gas disk calculation resulted in unreasonable small inner and outer radii. As we expand upon in Section \ref{sec:J0611}, the measured properties of this system require a more complex disk model than the flat disk model that we use in this paper.

\subsection{Brown Dwarf Template Comparison \label{sec: BD template comparison}}

We also compare the excess spectra from all of our systems with template spectra of field brown dwarfs. These templates were collected from the IRTF \citep{cushing2005-623} and SpeX libraries \citep{cruz2004-604,chiu2006-131,kirkpatrick2006-639,siegler2007-133,burgasser2008-681,looper2008-686,sheppard2009-137}, which include a total of 325 observed spectra of L and T dwarfs.

To find the best-match brown dwarf companion spectrum for each of the observed spectra, a $\chi^2$ type goodness-of-fit test was used as described in \cite{cushing2008-678}. The observed spectrum was first smoothed with a Gaussian filter as shown in Figure \ref{fig: all-spectra} to reduce noise and focus the fit on broad spectral features and the overall shape of the spectrum. For each template brown dwarf spectrum, a goodness-of-fit statistic \textit{G} was computed as described by Equation 1 of \cite{cushing2008-678}. This statistic \textit{G} was minimized with respect to a scaling factor \textit{C} multiplied to the flux of each template to match it to the flux level of the observed spectrum.

To estimate the most likely range of companion spectral types for each system, we ran our comparison algorithm on all the L and T dwarfs in the IRTF and SpeX libraries. For each spectral type, we calculated the mean \textit{G} value among the template spectra. We then found the lower 32\% percentile ($\mathrm{1\sigma}$) of average \textit{G} values among the spectral types, which we report as the likely range of companion spectral types for a given system in Table \ref{tab:MCMC_table}. Since Gaia~J0603+4518 had both \textit{MOSFIRE} and \textit{GNIRS} data, we report two separate ranges of companions that fit the data from each instrument. The \textit{MOSFIRE} spectra of Gaia~J0052+4505 and Gaia~J0603+4518 can be best fit with an L dwarf. The \textit{GNIRS} spectrum of Gaia~J0603+4518 can be best fit with a T dwarf. We discuss this discrepancy between the Gaia~J0603+4518 \textit{MOSFIRE} and \textit{GNIRS} data in Section \ref{sec: J0603}. Even though the Gaia~J0723+6301 spectrum can be best fit by an L dwarf, the lower $\mathrm{1\sigma}$ \textit{G} value is significantly higher compared to the other two systems, as shown in Figure~\ref{fig: BD-G-plots}, pointing to the fact that this system's spectrum was not well fit by any of the potential companions. 

Interestingly, the excess flux of all the disk systems can also be fit with an L dwarf template. However, their goodness of fit G values in Figure~\ref{fig: disk-G-plots} are much bigger than the likely brown dwarf systems in Figure~\ref{fig: BD-G-plots}.

So far, the majority of unresolved white dwarf brown dwarf pairs have been identified via systematic searches of white dwarfs for an infrared excess (e.g. \citealt{debes11, steele2011-416, girven11}). These systems were then confirmed via radial velocity measurements (e.g. \citealt{steele13, Maxted2006}). The radial velocity measurements can only detect systems with a short period, and hence high radial velocity. It is also often difficult to obtain high precision radial velocity measurements because it is hard to determine the centroid of broad Balmer lines in white dwarfs' atmospheres. One must also only use short exposures to avoid smearing the fast moving lines which is often challenging for these faint systems. In the most irradiated binaries there is emission  seen from the brown dwarf atmosphere allowing measurements from both components (e.g. WD0137-349AB; \citealt{casewell2015-447}). Since the launch of $Kepler$ and now with ZTF there have been more systems discovered to be eclipsing. These short period systems have deep, total eclipses where the Jupiter sized brown dwarf completely obscures the Earth sized white dwarf (e.g. \citealt{Parsons17, casewell2020-497, vanRoestel2021}). However, both of these methods of discovering binaries can be time consuming. It is not feasible to obtain a $\sim$10~hr lightcurve (or longer) of every white dwarf, or multiple epochs of spectroscopy in the hope of finding a companion.

Characterizing the infrared excess around white dwarfs is a much more efficient method to detect new white dwarf brown dwarf pairs \citep[e.g.][]{mansergas2022-927}.

\section{Discussion}\label{discussion}

\subsection{Gaia J0006+2858}

Optical observations of Gaia J0006+2858 have shown a rich set of emission features from Ca II, Fe II, O I, and possibly Mg I. These emission lines are unique in the significant asymmetry of their maximum blue and red velocities \citep{melis2020-905}. In the infrared spectra, we identified emission lines from Mg I, Fe I, and Si I, all of which are new detections. We also observed unidentified emissions near 1.6784 \micron~and 1.6875 \micron, which were also seen in Gaia~J2100+2122 (Table \ref{tab:emission lines}). The FWZIs of the infrared lines listed in Table~\ref{tab:emission lines} are around 1400~km~s$^{-1}$, which are generally consistent with those in the optical. The strongest infrared emission line is Mg I 1.5029~$\mu$m, which has an EW of 24~{\AA}. This is even stronger than the Ca II 8542~{\AA} emission from \cite{melis2020-905}, which had an EW of 19.2~{\AA}, and is typically the strongest emission feature in white dwarf gas disks. However, the Mg I 1.5029~$\mu$m line may be blended with nearby Mg I 1.5044~\micron~and 1.5051 \micron~lines, which contribute to the large FWZI of 1973 km s$^{-1}$. 

In addition to the new metal species identified in this system, our MCMC disk fit was able to model this system's infrared excess well, with excess beginning in the H band as shown in Figure \ref{fig: dust_fits}. The median-fit disk model for this system extends from 20 $\mathrm{R_{WD}}$ to 64 $\mathrm{R_{WD}}$, with an inclination 66\textdegree. Our fit was able to constrain the inner edge of the disk particularly well, with the upper and lower $\mathrm{1\sigma}$ region of the posterior distribution being only 4 $\mathrm{R_{WD}}$. 

Gaia~J0006+2858 also has a heavily polluted atmosphere, including detection of Ca, Mg, and Si \citep{Rogers2022}. The atmospheric pollution and strong metal emission lines show that this white dwarf hosts a dust disk.

\subsection{Gaia J0052+4505}

We identified broad molecular absorption features in the near-infrared spectrum of this system, indicative of a brown dwarf companion. In the J band, we identified $\mathrm{H_2O}$ absorption longward of 1.33 \micron. In the K band, we observed the CO band head beginning at 2.3 \micron. Both of these features are prominent in the spectra of brown dwarf atmospheres. We show these in Figure \ref{fig: BD comparison}, along with field brown dwarf spectra as a comparison.

We found that a brown dwarf companion was the most likely source of infrared excess flux for this system, given all of our observed data. Our goodness of fit tests on the morphology of this system's spectrum found that the most likely companion is an L6.5 $\pm$ 4.0 type brown dwarf. We note that the H band photometry for this system does not match up with the best-fitting brown dwarf model in Figure \ref{fig: BD_fits}. In Section \ref{sec: J0603}, we demonstrate that spectral variability may be seen for the other white dwarf brown dwarf system, Gaia J0603+4518. Since our photometric observations could have been taken at different phases of the companion's orbit, it is possible that we are seeing variability in the photometry here, if this is a tight, possibly tidally locked, system \cite{lew2021-163}. 

Our analysis of the \textit{Keck/HIRES} optical spectrum also did not find evidence of atmospheric pollution on the white dwarf. This lack of accretion further supports a brown dwarf companion being the source of infrared excess in this system.

\subsection{WD 0145+234}

The infrared spectrum of WD~0145+234 only shows absorption lines from the white dwarf's atmosphere. We did not identify any emission or absorption features. This is not particularly surprising, as previous optical studies only reported weak Ca II emissions \citep{melis2020-905}. Most interestingly, this system was discovered to be going through an ongoing burst in the mid-infrared by \cite{wang2019-886}. The system had brightened by about 1.0 magnitude in the \textit{W1} and \textit{W2} bands within half a year, while remaining the same in optical photometry. This led to the conclusion that the outburst was most likely due to a planetesimal being tidally disrupted, adding new material to an existing quiescent disk. WD~0145+234 is also heavily polluted with a high calcium accretion rate (as shown in Table~\ref{tab:emission lines}).
 
 As shown in Figure~\ref{fig: dust_fits}, our best fit disk model cannot reproduce all the observed photometry points. The larger variability in the WISE bands suggests strong variability in the Spitzer and JHK bands as well, which complicates the analysis for this system because the JHK photometry and \textit{Spitzer} photometry were taken at different times.

We conclude the infrared excess around WD 0145+234 comes from a dust disk due to the presence of strong infrared variability, a featureless infrared spectrum, and a heavily polluted atmosphere.

\subsection{Gaia J0603+4518}\label{sec: J0603}

Like Gaia J0052+4505, we identified broad molecular absorption features in the near-infrared spectrum of this system, indicating the presence of a brown dwarf companion. We observed $\mathrm{H_2O}$ absorption beginning at 1.33~\micron~in our \textit{GNIRS} spectrum, as well as the CO band head beginning at 2.3~\micron~in the K band of our \textit{MOSFIRE} spectrum. We show these in Figure \ref{fig: BD comparison}, along with field brown dwarf spectra as a comparison.

We could not make a clear detection of the CO band head in the \textit{GNIRS} spectrum of this object. The spectral resolution of \textit{GNIRS} is lower than that of \textit{MOSFIRE}, making it harder to detect weak features. Alternatively, this may be due to variability from viewing the dayside and nightside of a tidally locked brown dwarf companion. Near-infrared studies of other irradiated brown dwarfs in white dwarf binaries have shown a high level of wavelength-dependent variability in the flux contribution from the brown dwarfs, especially with respect to regions of molecular absorption. For example, \cite{lew2021-163} found the 1.3 $\micron$ $\mathrm{H_2O}$ band to vary 10 times more than the rest of the J band, in a brown dwarf irradiated by a white dwarf. In another similar system, \cite{casewell2018-481} observed a significant amount of variability between dayside and nightside K band photometry, where the CO band head is located. The effective temperature of this white dwarf is comparable to WD~0137-349, which hosts an irradiated brown dwarf with observed variability in optical and near-infrared photometric studies \citep{casewell2015-447,zhou2022-163}. This further demonstrates the plausibility of Gaia~J0603+4518 hosting a similarly irradiated brown dwarf companion. Future observations with time-series infrared spectroscopy would give a better understanding of how the CO band head is affected by irradiation in a tight white dwarf brown dwarf binary system.

In addition, we observed some differences in the overall morphology between the \textit{MOSFIRE} and \textit{GNIRS} spectra. In our comparison tests with template brown dwarf spectra, as described in Section \ref{sec: BD template comparison}, these morphological differences resulted in slightly different best-fitting companions for the two spectra. Our tests found that the \textit{MOSFIRE} spectrum best followed an early to late L dwarf, while the \textit{GNIRS} spectrum best followed a late L to early T dwarf.

Our modeling of the observed \textit{MOSFIRE} and \textit{GNIRS} spectra of this system found a brown dwarf companion to be more likely than a debris disk as the source of infrared excess, which begins in the J band. The goodness of fit test for the \textit{MOSFIRE} data estimates this companion to be an L3.5 $\pm$ 3.5 type brown dwarf, while the \textit{GNIRS} spectrum favored a T2.5 $\pm$ 2.5. Figure \ref{fig: BD-G-plots} illustrates the difference in the goodness of fit tests for the two instruments. We again believe that these differences in the morphology of the spectrum between the two instruments for this system is due to observing the brown dwarf companion in different phases of its orbit, as the two epochs of data were taken about a year apart.

Our analysis of the \textit{Keck/HIRES} optical spectrum did not find evidence of atmospheric pollution on the white dwarf.  This lack of accretion further supports the scenario of brown dwarf companion being the source of infrared excess in this system. Gaia~J0603+4518 appears as one single object in the GNIRS and MOSFIRE data, which means the companion must be within 0{\farcs}7 (projected separation of 42~AU) of the white dwarf.

\subsection{Gaia J0611--6931 \label{sec:J0611}}

We identified strong emission lines from Mg I, Fe I, and Si I in the \textit{Flamingos-2} spectrum of Gaia~J0611--6931. Previously, emissions features from O I, Na I, Mg I, Si I, Ca II, and Fe II have been reported in the optical spectra \citep{dennihy2020-905,melis2020-905}. These infrared Mg I, Fe I, and Si I lines are much stronger than those in the optical. The FWZIs reported in Table~\ref{tab:emission lines} are also larger than FWZIs of 1400~km s$^{-1}$ reported in the optical, likely due to blending of several lines. With six different species, Gaia~J0611--6931 has the most variety of metal emissions.

In addition to this rich emission spectrum, Gaia~J0611--6931 has one of the brightest known infrared excesses from a white dwarf, with a fractional luminosity of 5.16\% \citep{dennihy2020-905}. While our best fitting disk model performed better than the best brown dwarf companion, the level of excess from this system could not be reproduced by our model. As shown in Figure~\ref{fig: dust_fits}, even a wide face-on disk fails to produce the strong infrared excess. Relaxing the physical constraints on the disk's inner and outer disk radius did not improve the fit. \cite{dennihy2020-905} suggested that the strength of this excess would require a model with multiple dusty components rather than a flat dust disk model. Simulations show that the disk could retain a large scale height under collisional cascade \citep{Kenyon2017,Ballering2022}.

High resolution optical spectroscopy has revealed the presence of Ca, O, Mg, Si, and Fe in the white dwarf atmosphere \citep{Rogers2022}. The presence of a strong infrared excess, the detection of multiple metal emission features, and a heavily polluted atmosphere provide convincing evidence that Gaia~J0611--6931 hosts a dust disk.

\subsection{Gaia J0723+6301}
    
We found the near-infrared spectrum of this system to be consistent with a white dwarf until the K band where excess flux begins (see Figure~\ref{fig: BD_fits}). We did not identify any emission features or broad absorption features indicating either a debris disk or brown dwarf companion.

Our analysis of the \textit{Keck/HIRES} spectrum of places an upper limit of the calcium mass accretion rate of 7.5 $\times$ 10$^5$ g s$^{-1}$. Assuming the calcium mass fraction is 1.62\%, taken from bulk Earth \citep{allegre2001-185}, in the accreting material, the upper limit to the total mass accretion rate is 4.6 $\times$ 10$^7$ g s$^{-1}$, which is very low compared to other dusty white dwarfs \citep{xu2019-158}. If the infrared excess comes from a dust disk, Gaia~J0723+6301 would have the lowest mass accretion rate among all the dusty white dwarfs.

We also checked the \textit{Spitzer/IRAC} images of Gaia~J0723+6301 for a potential background galaxy causing the infrared excess. The white dwarf is well detected in both IRAC-1 and IRAC-2 and it appears to be isolated. Aperture photometry and PRF photometry returns the same flux level for this system \citep{lai2021-920}. The GNIRS acquisition image also shows that there are no objects beyond 1{\farcs}0 (projected separation 137.9~AU) of the white dwarf. However, we cannot completely rule out the possibility that there is a background galaxy along the line of sight and only emits in the IRAC bands. In that scenario, the infrared excess around Gaia J0723+6301 is spurious.

From the literature, there is one system in a similar situation as Gaia~J0723+6301 -- WD~2328+107, which has a subtle infrared excess \citep{Rocchetto2015} but no atmospheric pollution from {\it Hubble} observations \citep{Wilson2019}. There are no infrared spectroscopic observations on WD~2328+107, which would be very useful in characterizing the origin of the infrared excess.

\subsection{Gaia J2100+2122}

The optical spectrum of Gaia~J2100+2122 is dominated by Fe II emission lines, as well as Ca II and O I  \citep{dennihy2020-905,melis2020-905}. An interesting characteristic of the system is that the emission line strength varied dramatically over the course of a few months, likely due to ongoing gas production or excitation. We have tentative identifications of Mg I, Fe I, and Si I in the \textit{MOSFIRE} spectrum of this system. As listed in Table~\ref{tab:emission lines}, the Si I lines are much stronger compared to Mg I and Fe I lines, which is the opposite of Gaia~J0611--6931 and Gaia~J0006+2858. This may be explained by different silicon to magnesium ratio in the gas debris. We also observed unidentified emissions near 1.6784 \micron~and 1.6875 \micron, also seen in Gaia~J0006+2858.

Given the observed data, we found the most likely scenario for the source of infrared excess in this system to be a dusty debris disk. Fitting the infrared excess flux as a flat disk, we constrained the disk inner radius to be near 33 R$\mathrm{_{WD}}$, with an outer radius likely extending to 57 R$\mathrm{_{WD}}$. We show this disk model in Figure \ref{fig: dust_fits}.

The atmosphere of Gaia~J2100+2122 is polluted with Ca, Mg, Fe, and Si \citep{Rogers2022}. The detection of variable metal emissions, a nearly featureless infrared spectrum, and a heavily polluted atmosphere suggest that Gaia~J2100+2122 hosts a dust disk.

\section{Conclusion}\label{sec:conclusion}

We present near-infrared spectroscopy of seven infrared excess white dwarf systems. With these new data, we are able to determine the source of infrared excess in six of the white dwarf systems. For the first time, we identified infrared emission features (Mg I, Fe I, and Si I) from the gaseous component of a debris disk around a white dwarf. We also clearly observed broad molecular absorption features from the atmospheres of two unresolved white dwarf brown dwarf pairs.

We showed, with a high level of confidence, that the sources of infrared excess in four of the white dwarf systems---Gaia~J0006+2858, WD~0145+234, Gaia~J0611--6931, and Gaia~J2100+2122---are dusty debris disks. We detected emissions from Mg I, Fe I, and Si I in the spectra of Gaia~J0611+6931, Gaia~J0006+2858, and Gaia~J2100+2122 which originate from a gaseous component of their debris disks. We also found the best-fitting parameters for a flat dust disk model for each of these systems, which fit the observed data for Gaia~J0006+2858 and Gaia~J2100+2122 particularly well. Using the best-fitting inclination range from these modeled disks along with optical gas emissions reported in \cite{melis2020-905}, we were able to calculate the inner and outer radii of emitting gas in the four disk systems. We found that the inner and outer radii of the gas disks were consistent with the best-fitting location of the dust disks for Gaia~J0006+2858, WD~0145+234, and Gaia~J2100+2122, supporting theories about the formation of the gas and dust disks. The infrared excess observed around Gaia~J0611--6931 is so strong that the disk cannot be flat. This system will require more complex models of gas and dust debris components in order to reproduce its measured properties.

We found strong evidence for the presence of unresolved brown dwarf companions as the source of infrared excess in two of our systems---Gaia~J0052+4505 and Gaia~J0603+4518. These systems both showed broad absorption features attributed to molecules in the atmospheres of brown dwarfs. When we compared the observed spectra for each of these systems to a large sample of brown dwarf spectra, we determined the best-fitting companions for Gaia~J0052+4505 and Gaia~J0603+4518 were L dwarfs.

The source of infrared excess in Gaia~J0723+6301 remains a mystery. There is a lack of detectable metal pollution from the Keck/HIRES data and there are no signs of molecular absorption features that would support the presence of a brown dwarf companion. Thus, we do not have enough evidence to discern the most likely source of excess in this system. 

These seven systems come from the larger sample of infrared excess white dwarfs reported in \cite{lai2021-920}. Future follow-up spectroscopy in the infrared of these systems would lead to a much better understanding of their occurrence rates, the unique nature of individual systems, and their properties as a population. Observations in even longer wavelengths, such as with the {\it James Webb Space Telescope}, will much better constrain the parameters of these systems.

\vspace{5mm}
{\it Data Availability}

\begin{figure*}[tbh]
    \centering    
    \includegraphics[height=0.23\textheight]{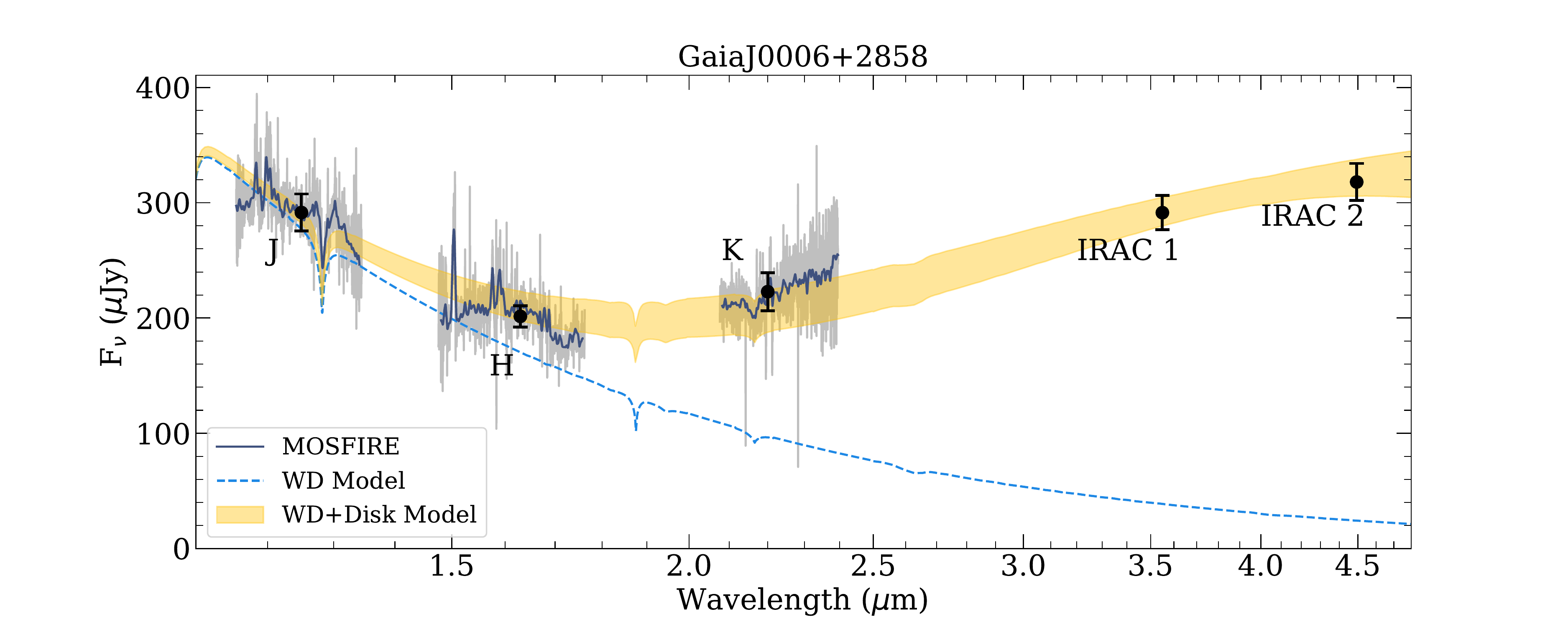}
    \includegraphics[height=0.23\textheight]{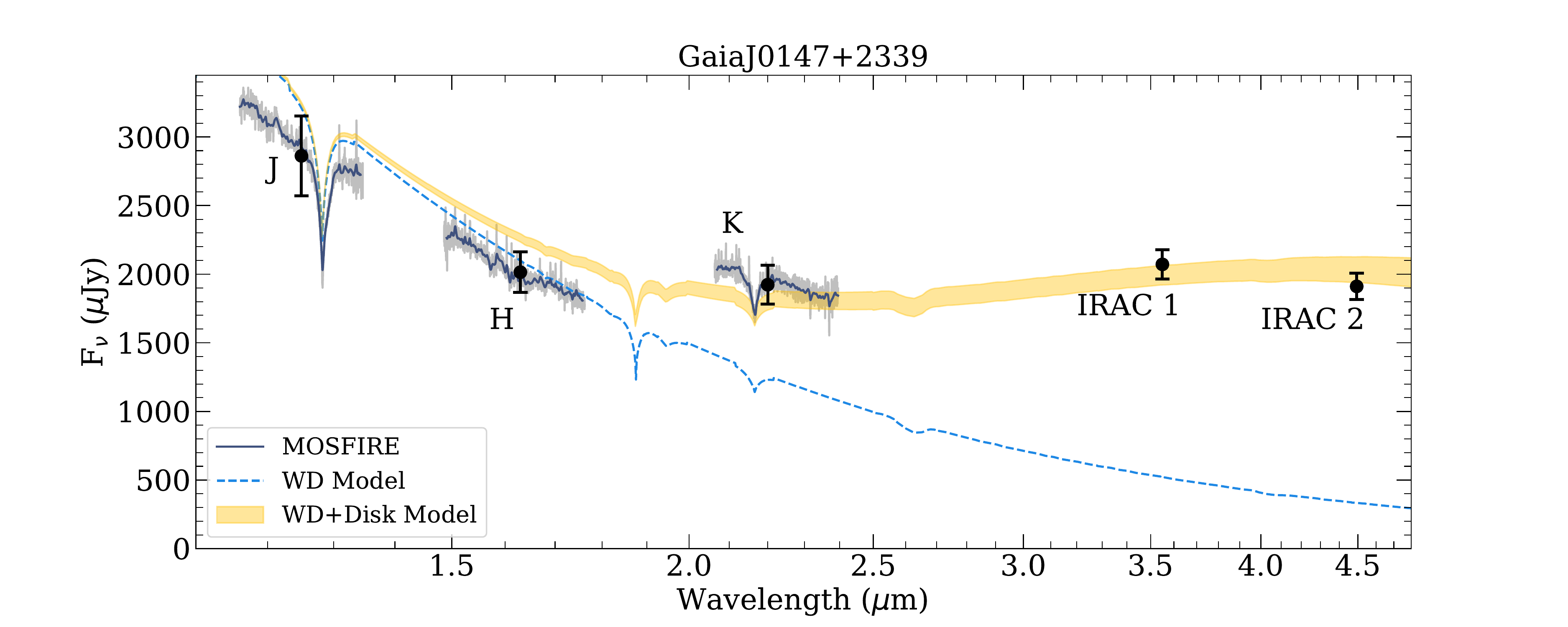}
    \includegraphics[height=0.23\textheight]{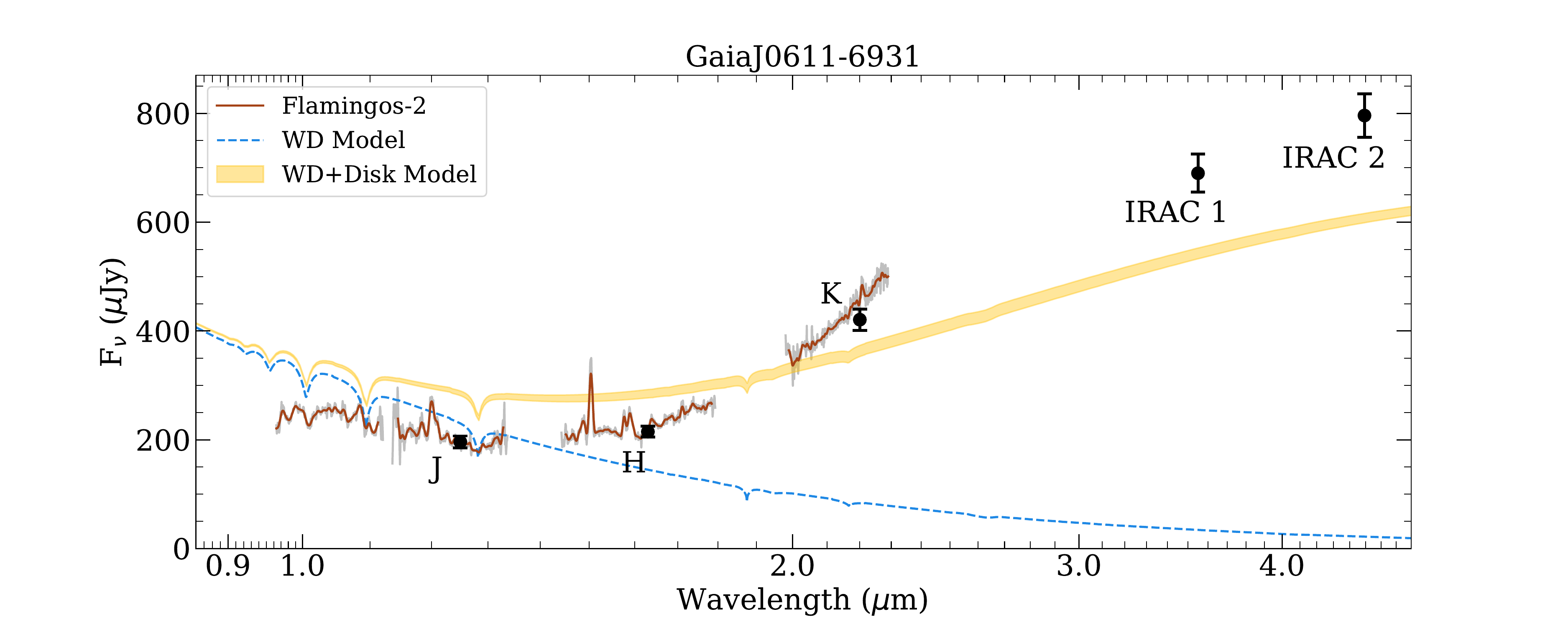}
    \includegraphics[height=0.23\textheight]{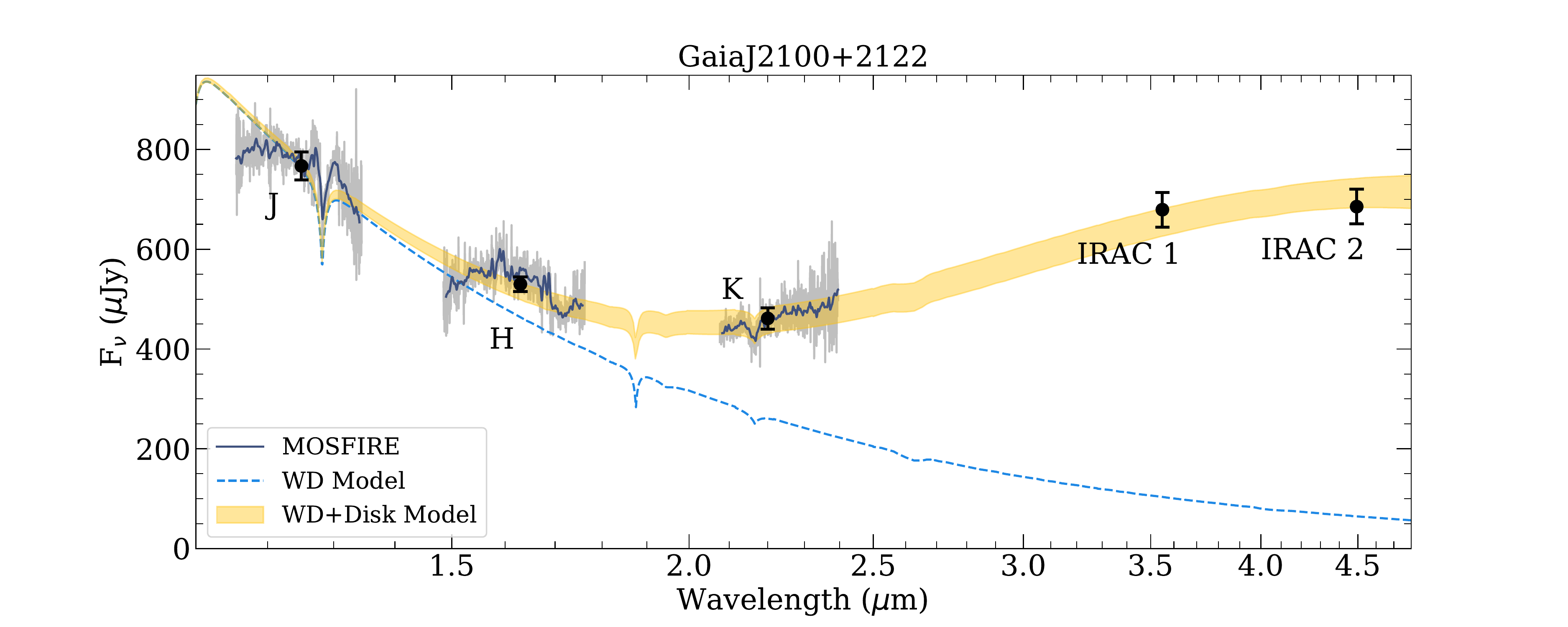}
    \caption{The four dust disk systems, plotted with their median-fit dust model and $\mathrm{1\sigma}$ uncertainty region, along with a white dwarf model spectrum. The disk parameters are listed in Table~\ref{tab:MCMC_table}.
    \label{fig: dust_fits}
    }
\end{figure*}

\begin{figure*}[tbh]
    \centering
    \includegraphics[height=0.3\textheight]{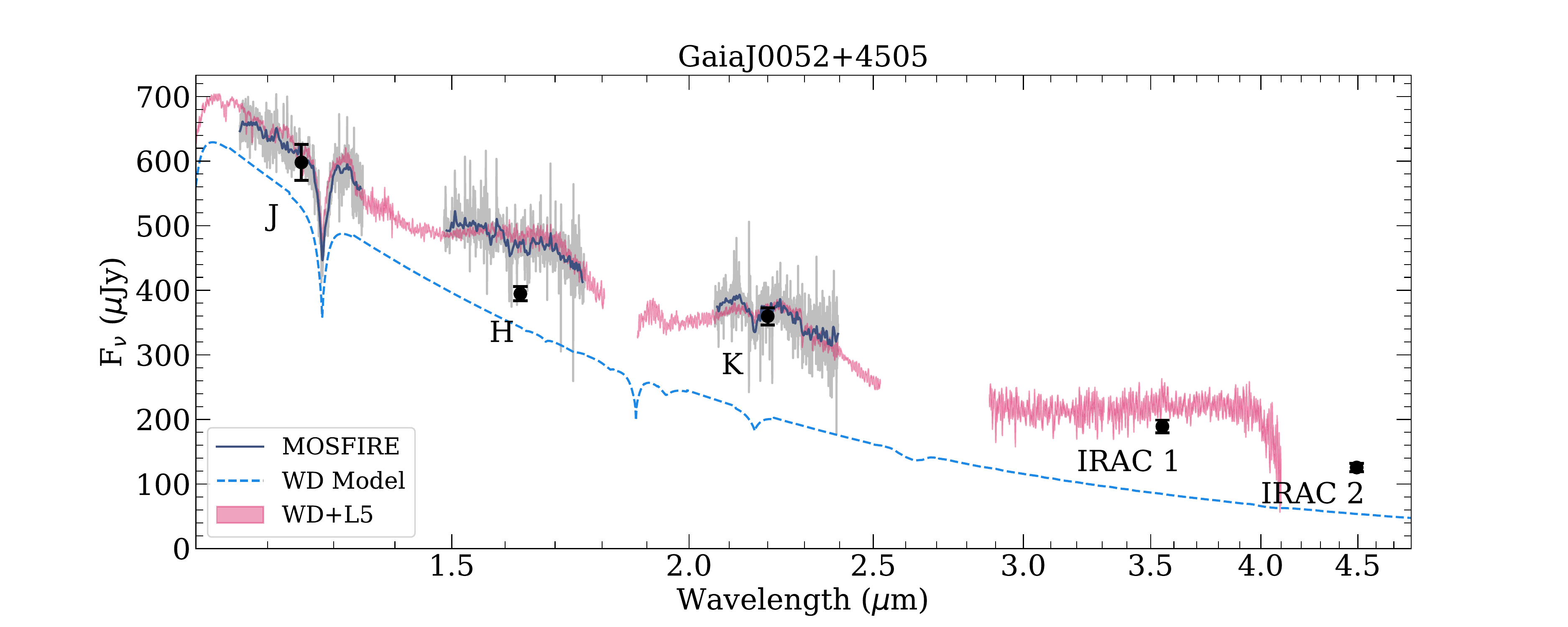}
    \includegraphics[height=0.3\textheight]{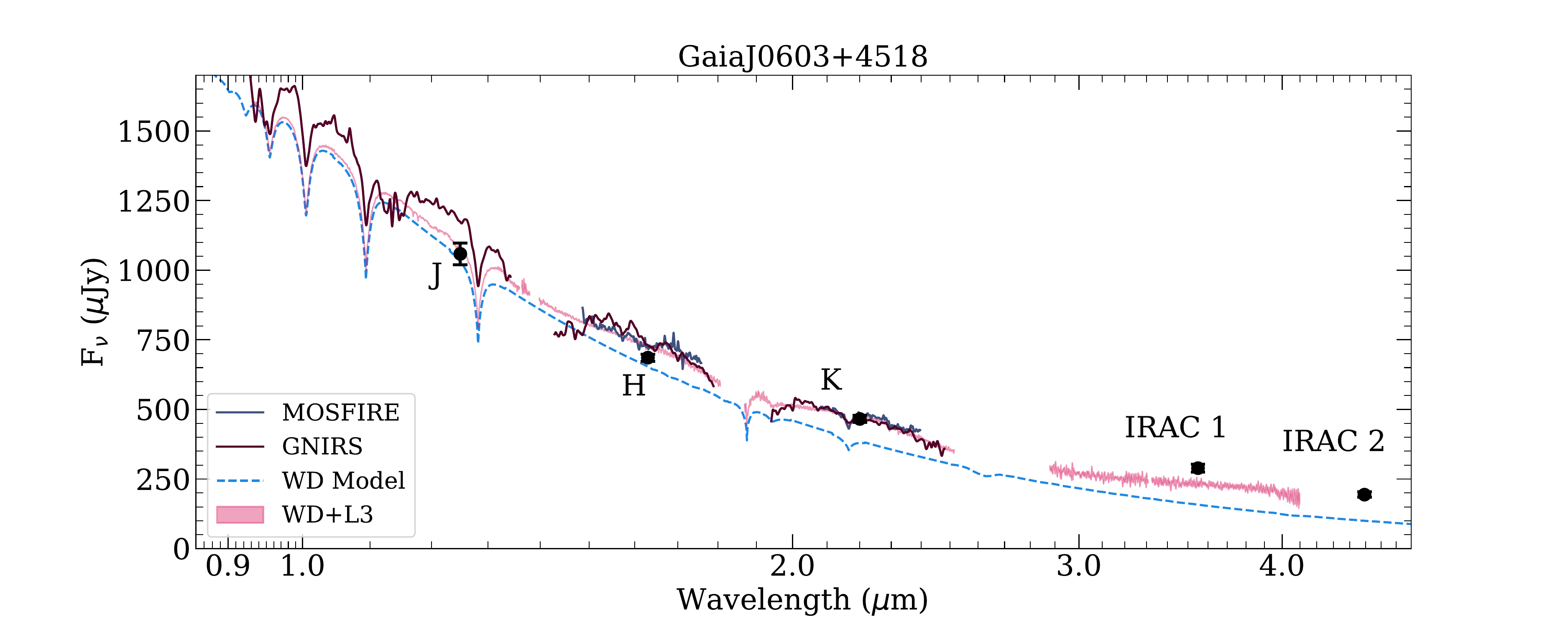}
    \includegraphics[height=0.3\textheight]{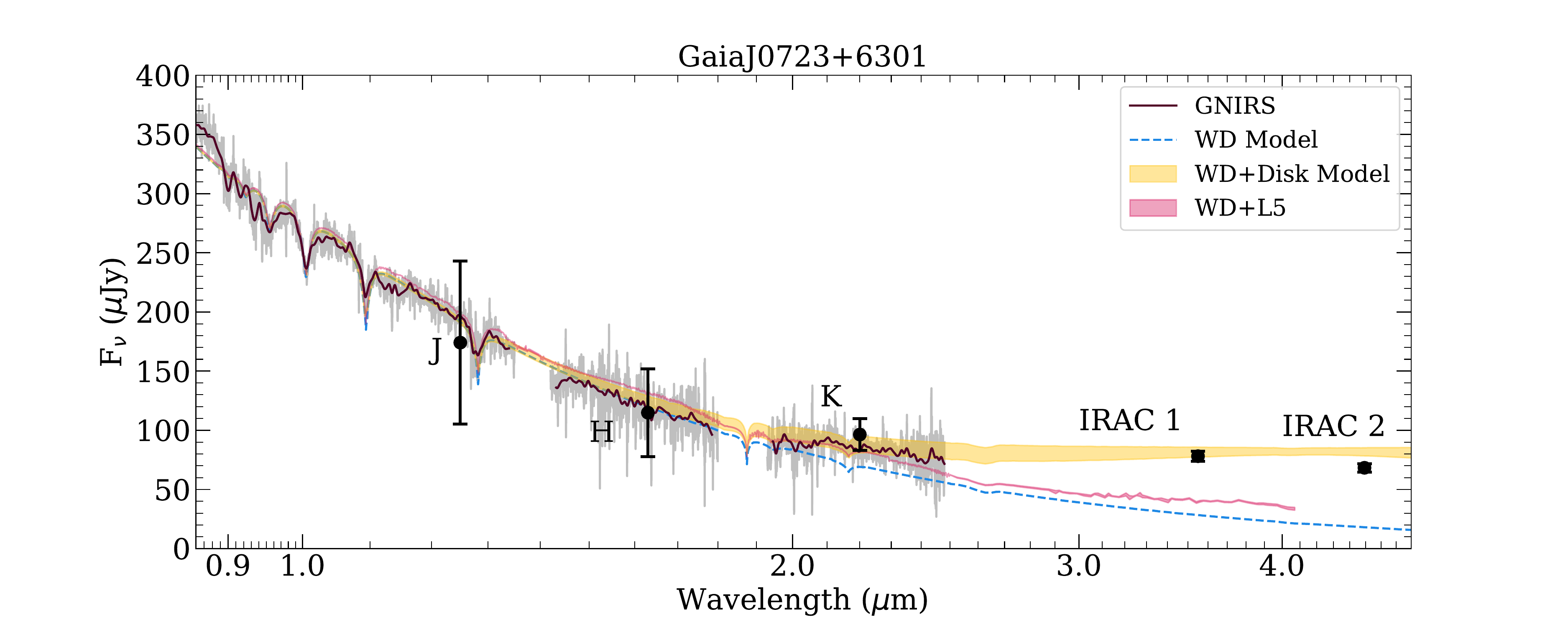}
    \caption{The two brown dwarf companion systems (Gaia J0052+4505, Gaia J0603+4518) and Gaia~J0723+6301, plotted with their best-fit brown dwarf companion template spectrum. We also show the best fit disk model for Gaia~J0723+6301. 
    \label{fig: BD_fits}
    }
\end{figure*}

The reduced \textit{MOSFIRE}, \textit{GNIRS}, \textit{Flamingos-2} infrared spectroscopy described in Section \ref{sec:obs and red} and shown in Figures \ref{fig: all-spectra},\ref{fig: dust_fits}, and \ref{fig: BD_fits} are available in machine readable format accompanying paper. We report the wavelength axis in $\micron$, and the flux axis in $\mathrm{\mu Jy}$.

\vspace{5mm}
{\it Acknowledgements.} This research made use of {\ttfamily{PypeIt}},\footnote{\url{https://pypeit.readthedocs.io/en/latest/}}
a Python package for semi-automated reduction of astronomical slit-based spectroscopy
\citep{pypeit:joss_pub, pypeit:zenodo}.

We would like to thank Samuel Lai for his help with our questions about the photometry in \cite{lai2021-920}. We would also like to thank the Gemini-North staff for creating such a welcoming environment to work in.

This work is based on observations obtained at the international Gemini Observatory, a program of NSF's NOIRLab, which is managed by the Association of Universities for Research in Astronomy (AURA) under a cooperative agreement with the National Science Foundation on behalf of the Gemini Observatory partnership: the National Science Foundation (United States), National Research Council (Canada), Agencia Nacional de Investigaci\'{o}n y Desarrollo (Chile), Ministerio de Ciencia, Tecnolog\'{i}a e Innovaci\'{o}n (Argentina), Minist\'{e}rio da Ci{\^e}ncia, Tecnologia, Inova\c{c}\~{o}es e Comunica\c{c}\~{o}es (Brazil), and Korea Astronomy and Space Science Institute (Republic of Korea). The data is processed using the Gemini IRAF package. This work was enabled by observations made from the Gemini North telescope, located within the Maunakea Science Reserve and adjacent to the summit of Maunakea. We are grateful for the privilege of observing the Universe from a place that is unique in both its astronomical quality and its cultural significance.

The data presented herein were obtained at the W. M. Keck Observatory, which is operated as a scientific partnership among the California Institute of Technology, the University of California and the National Aeronautics and Space Administration. The Observatory was made possible by the generous financial support of the W. M. Keck Foundation. This work is partly supported by the Heising Simons foundation.

This research has benefitted from the SpeX Prism Spectral Libraries, maintained by Adam Burgasser at http://www.browndwarfs.org/spexprism.

\facilities{Gemini:Gillett (GNIRS, Flamingos2), Keck: I (MOSFIRE, HIRES)}

\software{Astropy \citep{astropy2022}, emcee \citep{foreman-mackey2013-125}}

\clearpage
\appendix
\setcounter{figure}{0} 
\renewcommand\thefigure{A\arabic{figure}} 
\renewcommand*{\theHfigure}{A\arabic{figure}}
The corner plots for the MCMC modeling described in section~\ref{sec: disk model fitting} is shown in Figures~\ref{fig: disk-corners} and \ref{fig: BD-corners}. The goodness of fit statistic G described in section~\ref{sec: BD template comparison} is shown in Figure~\ref{fig: disk-G-plots} and \ref{fig: BD-G-plots}.

\begin{figure*}[tbh]
    \centering
    \includegraphics[width=0.49\textwidth]{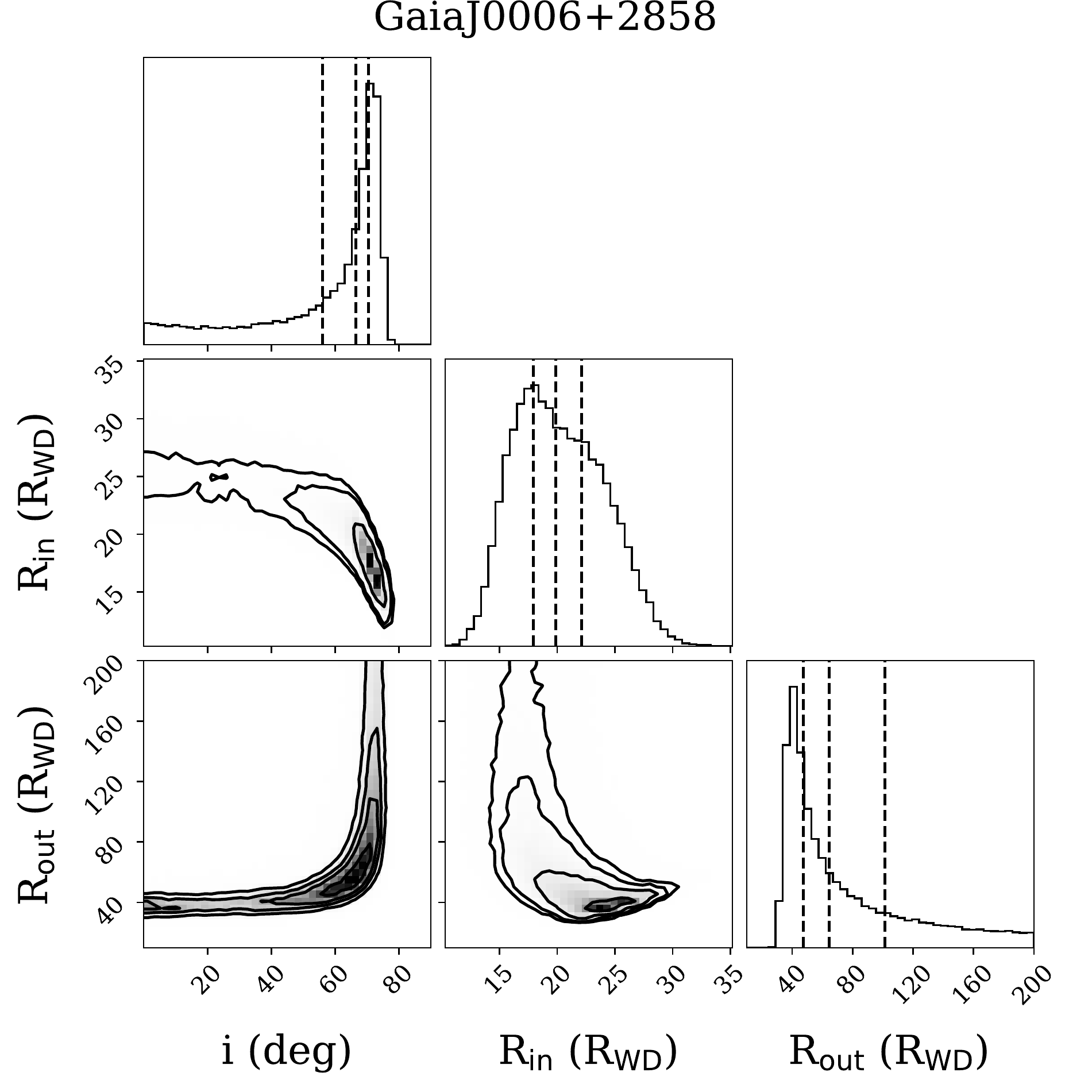}
    \includegraphics[width=0.49\textwidth]{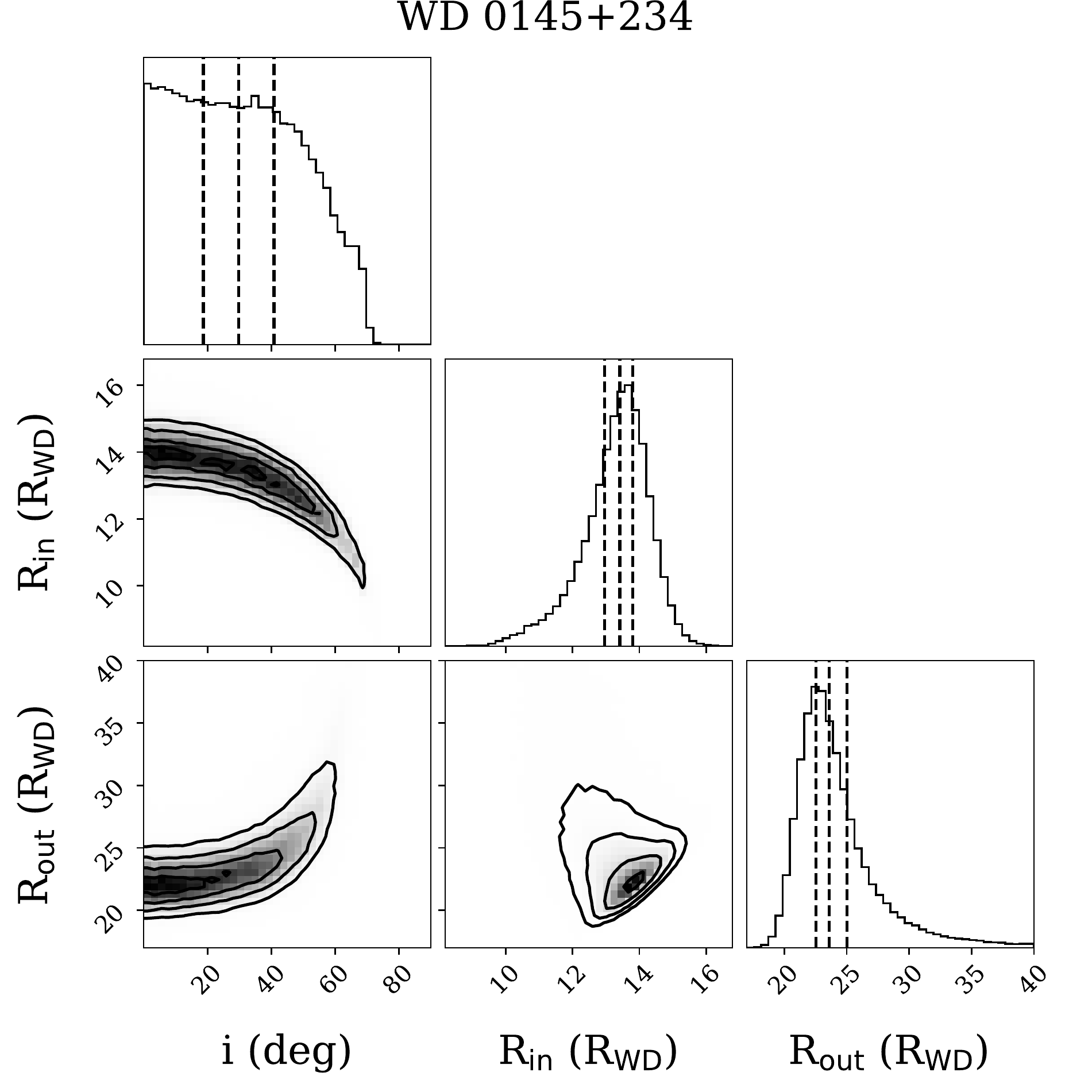}
    \includegraphics[width=0.49\textwidth]{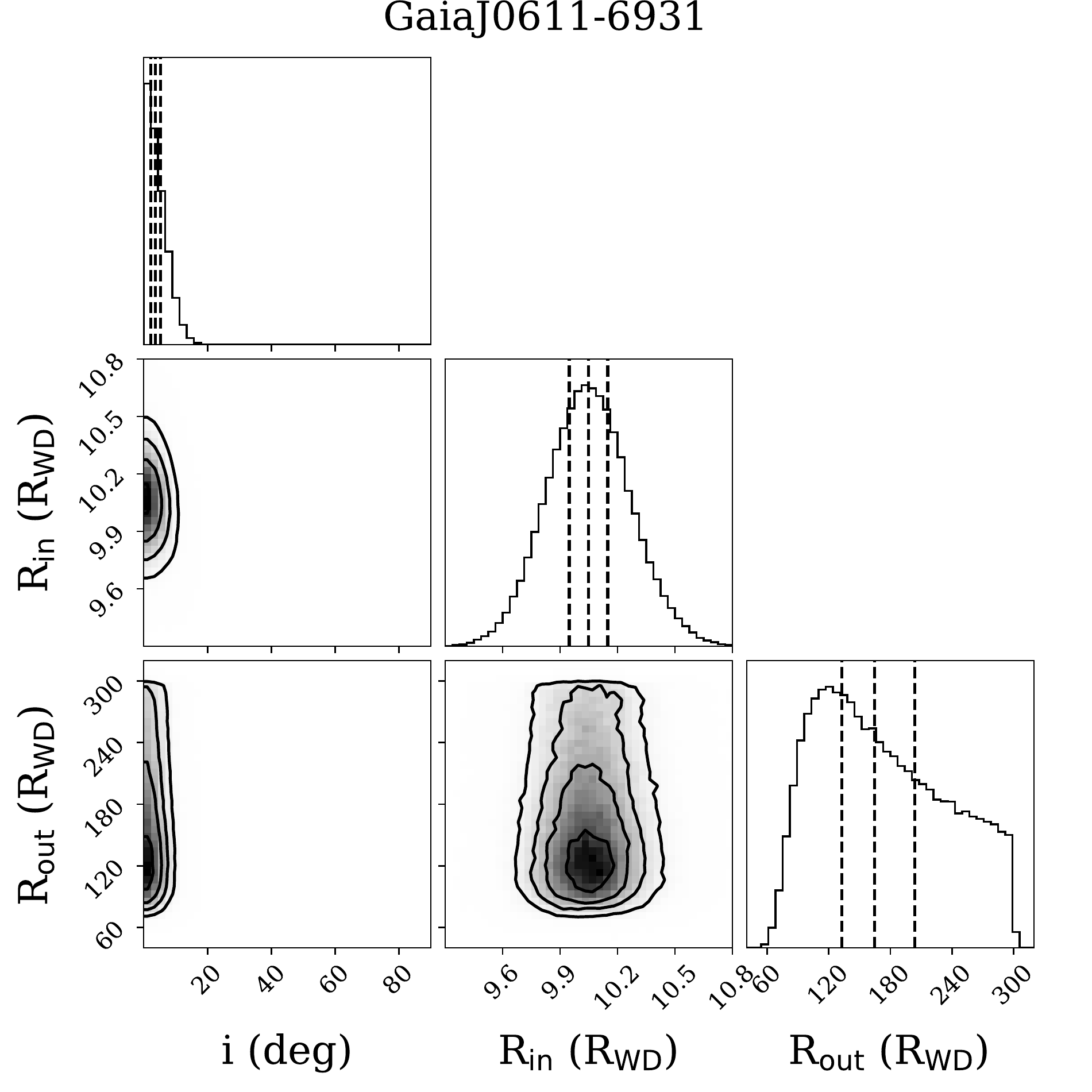}
    \includegraphics[width=0.49\textwidth]{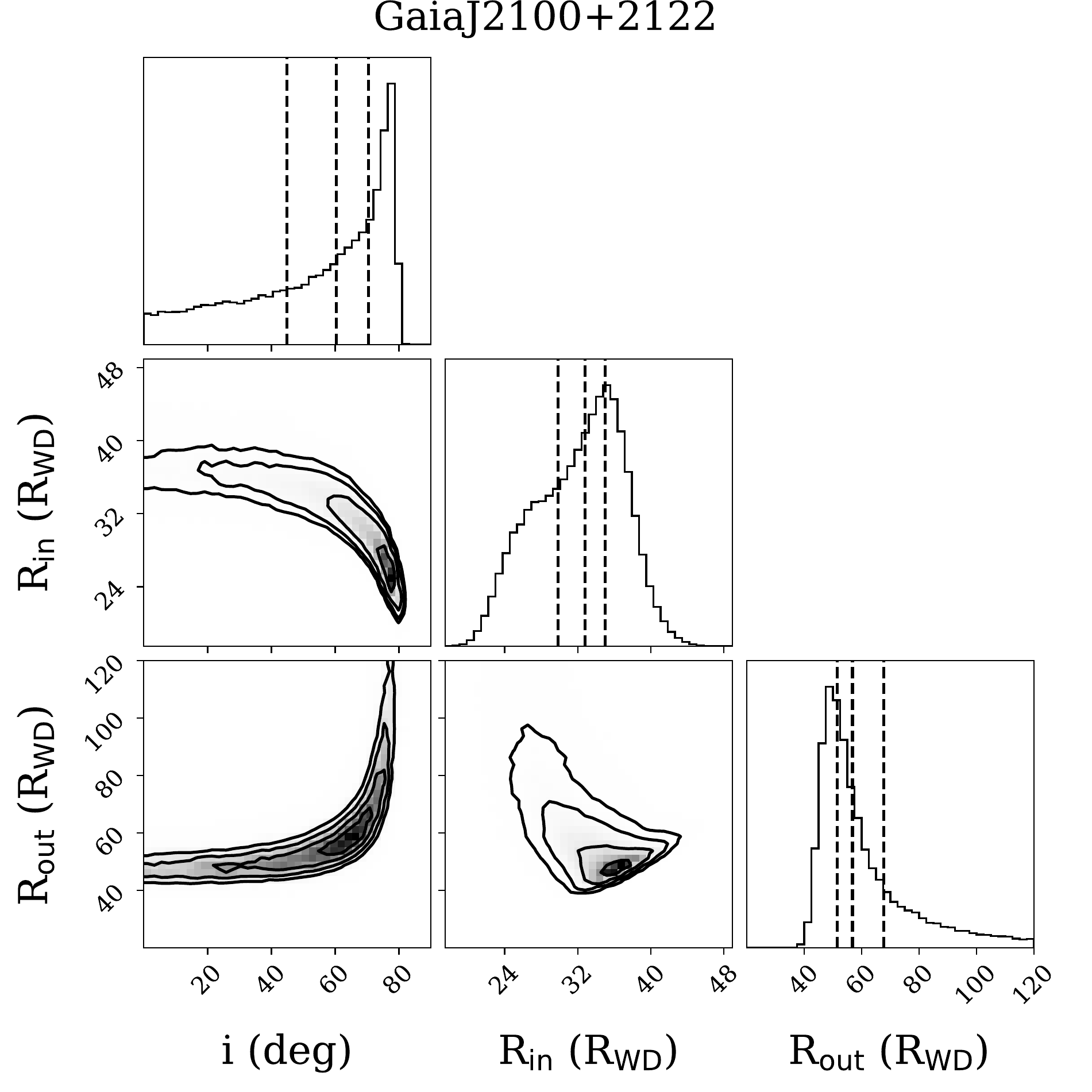}
    \caption{Corner plots from the MCMC fits of the disk systems. The dashed lines mark the median values and 1$\sigma$ uncertainties. For each system, the inner radius (R$_\mathrm{in}$) is generally well constrained.
    \label{fig: disk-corners}
    }
\end{figure*} 

\begin{figure*}[tbh]
    \centering
    \includegraphics[width=0.49\textwidth]{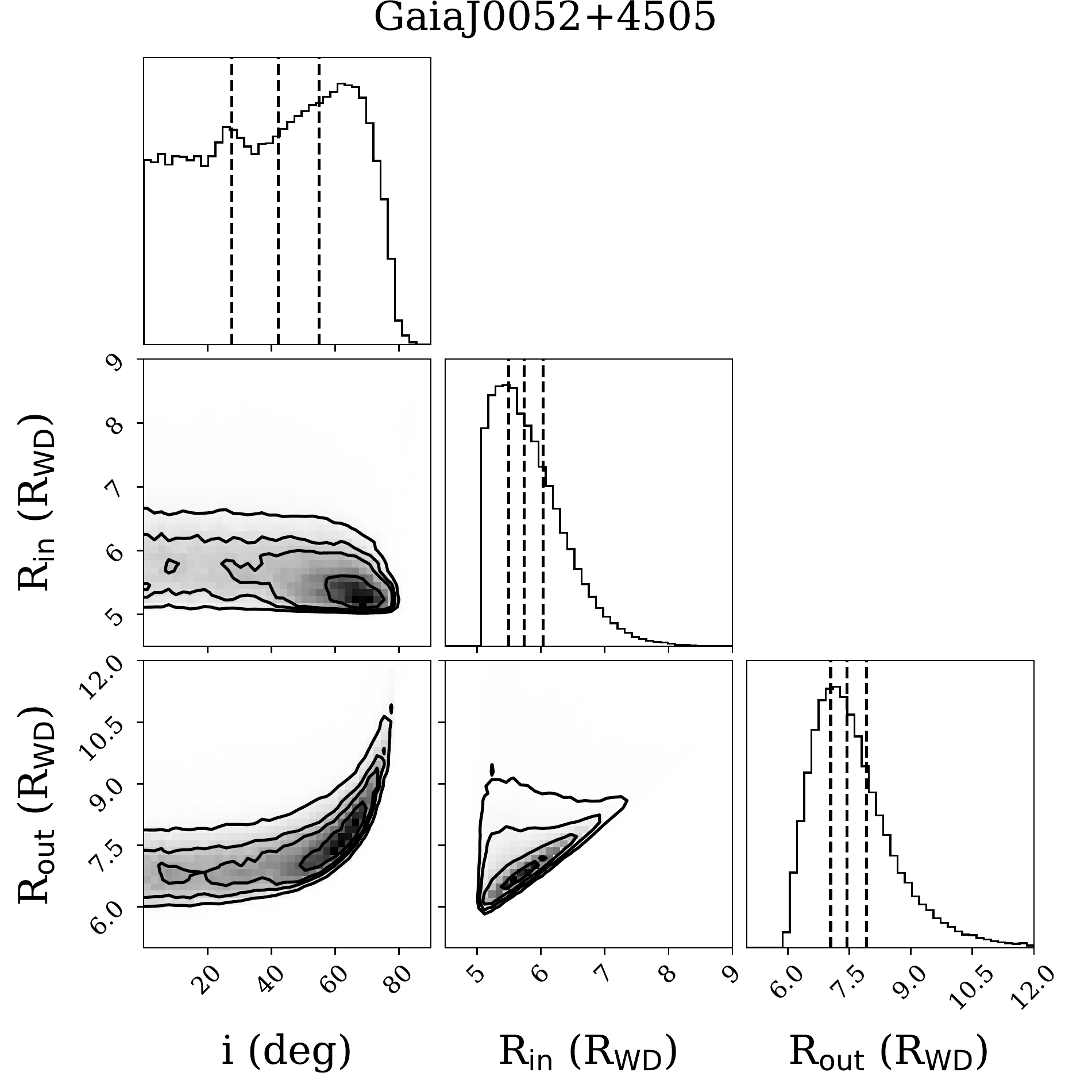}
    \includegraphics[width=0.49\textwidth]{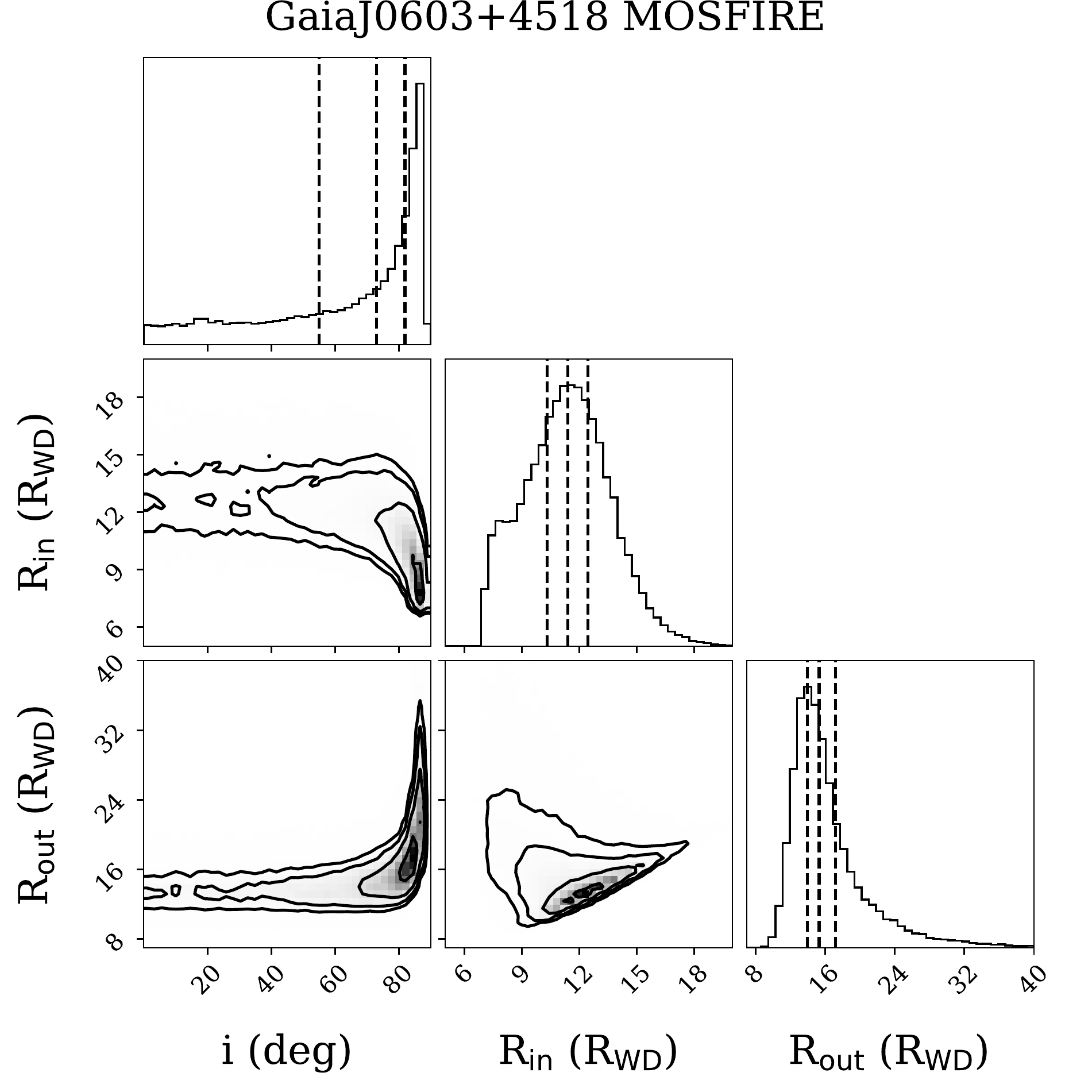}
    \includegraphics[width=0.49\textwidth]{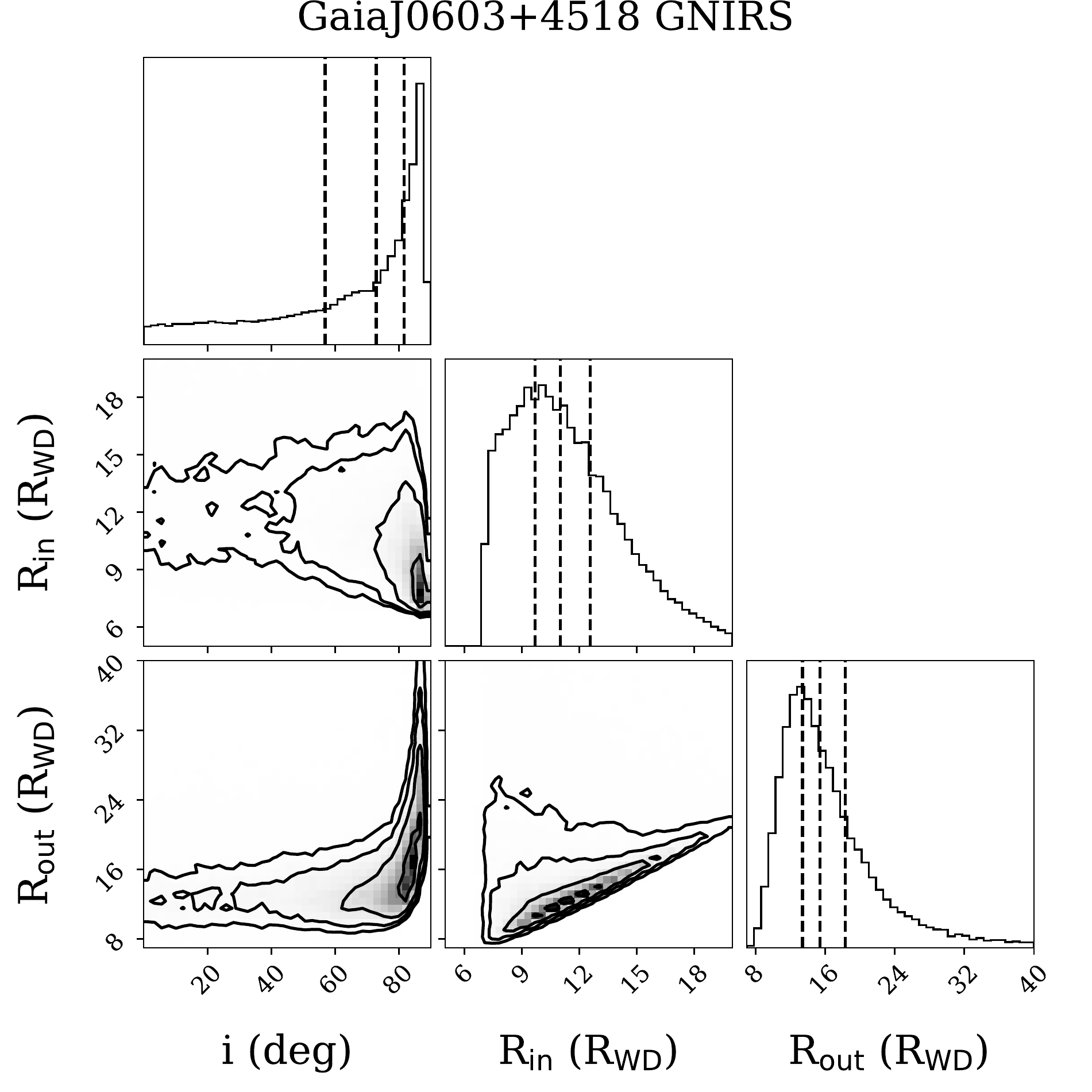}
    \includegraphics[width=0.49\textwidth]{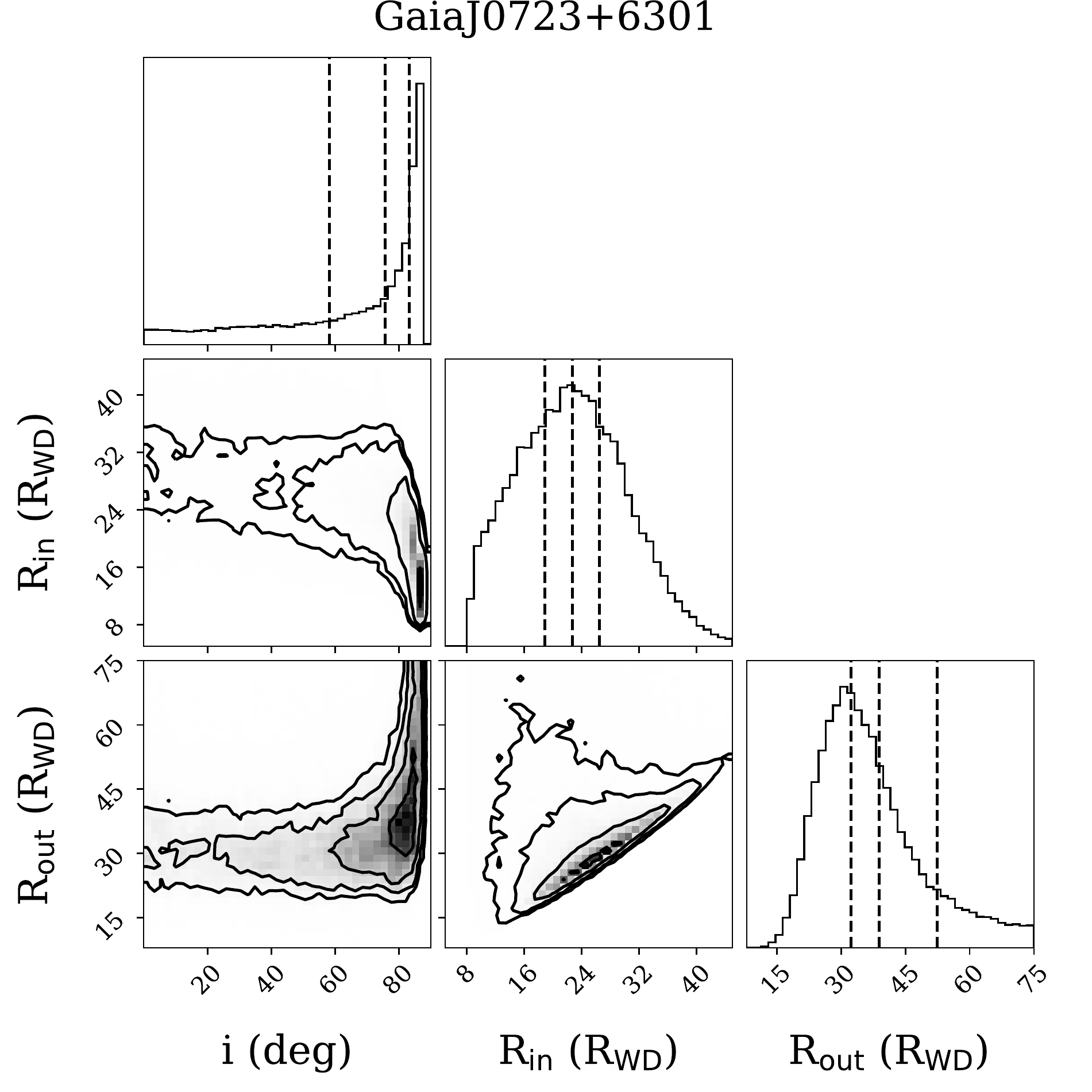}
    \caption{Similar to Figure~\ref{fig: disk-corners} for the three remaining systems. Compared to the disk systems shown in Figure~\ref{fig: disk-corners}, these infrared excesses can be fit by narrower disks due to the smaller emitting areas.
    \label{fig: BD-corners}
    }
\end{figure*} 

\begin{figure*}[tbh]
    \centering
    \includegraphics[width=0.65\textwidth]{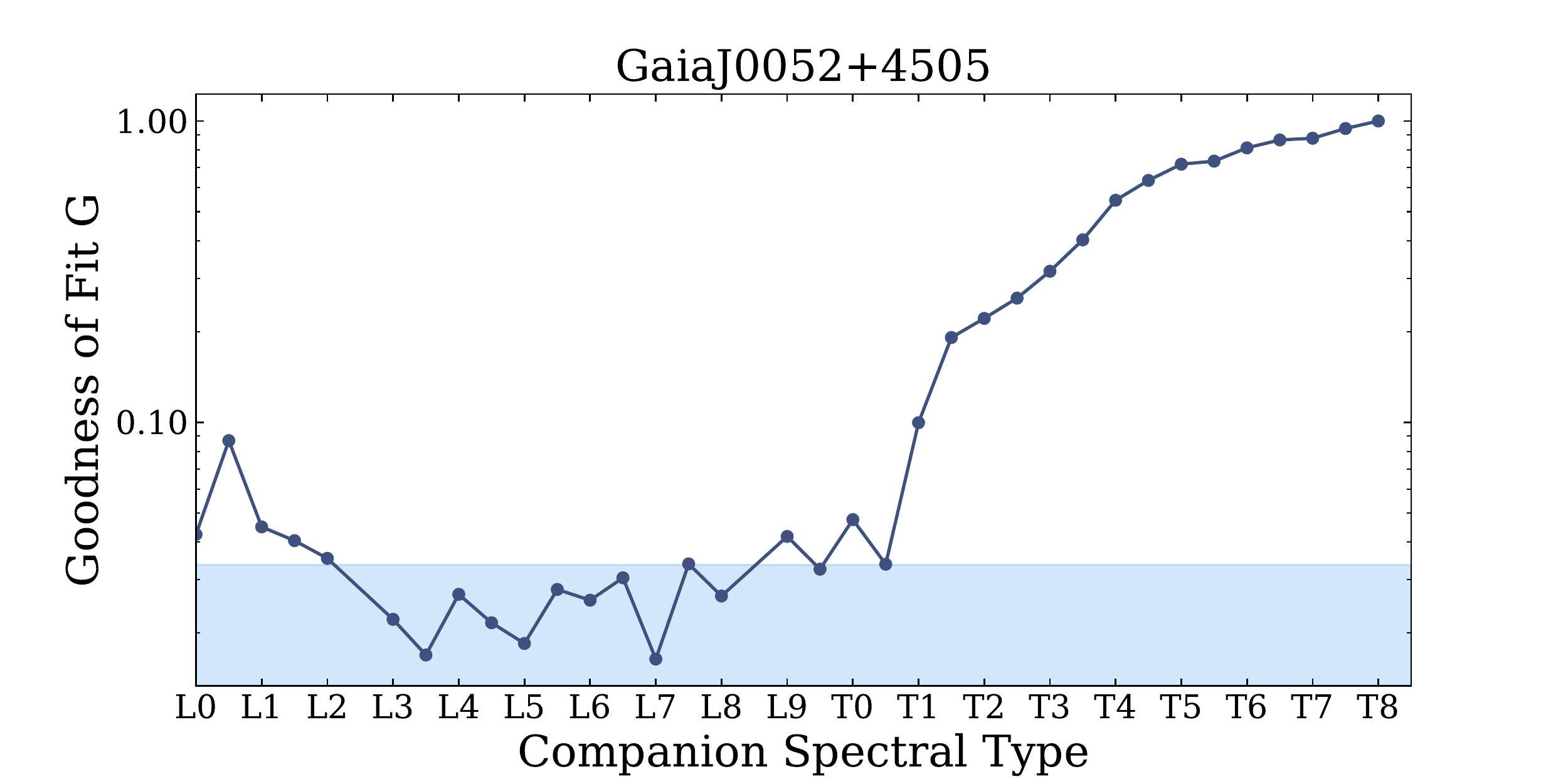}
    \includegraphics[width=0.65\textwidth]{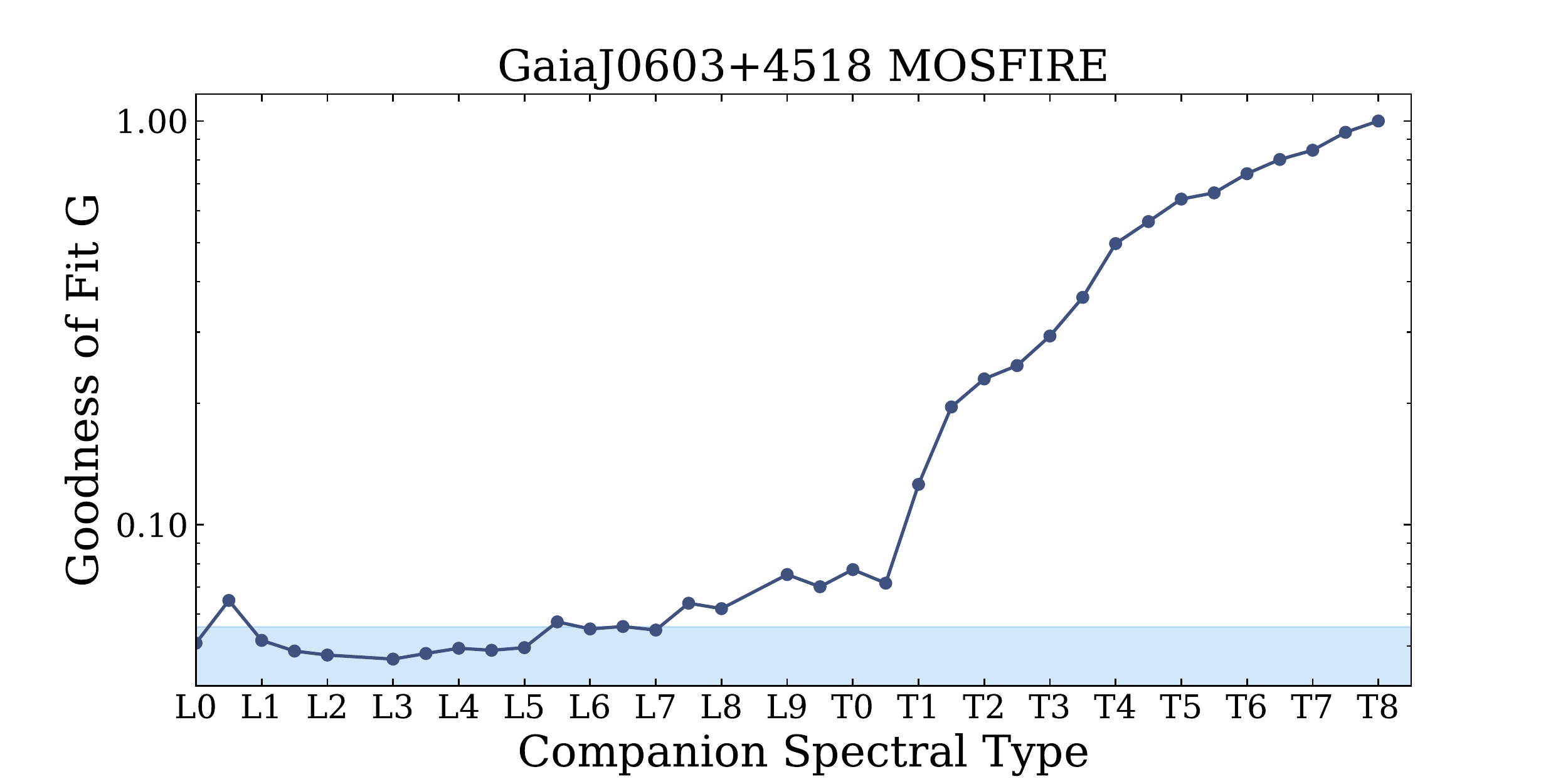}
    \includegraphics[width=0.65\textwidth]{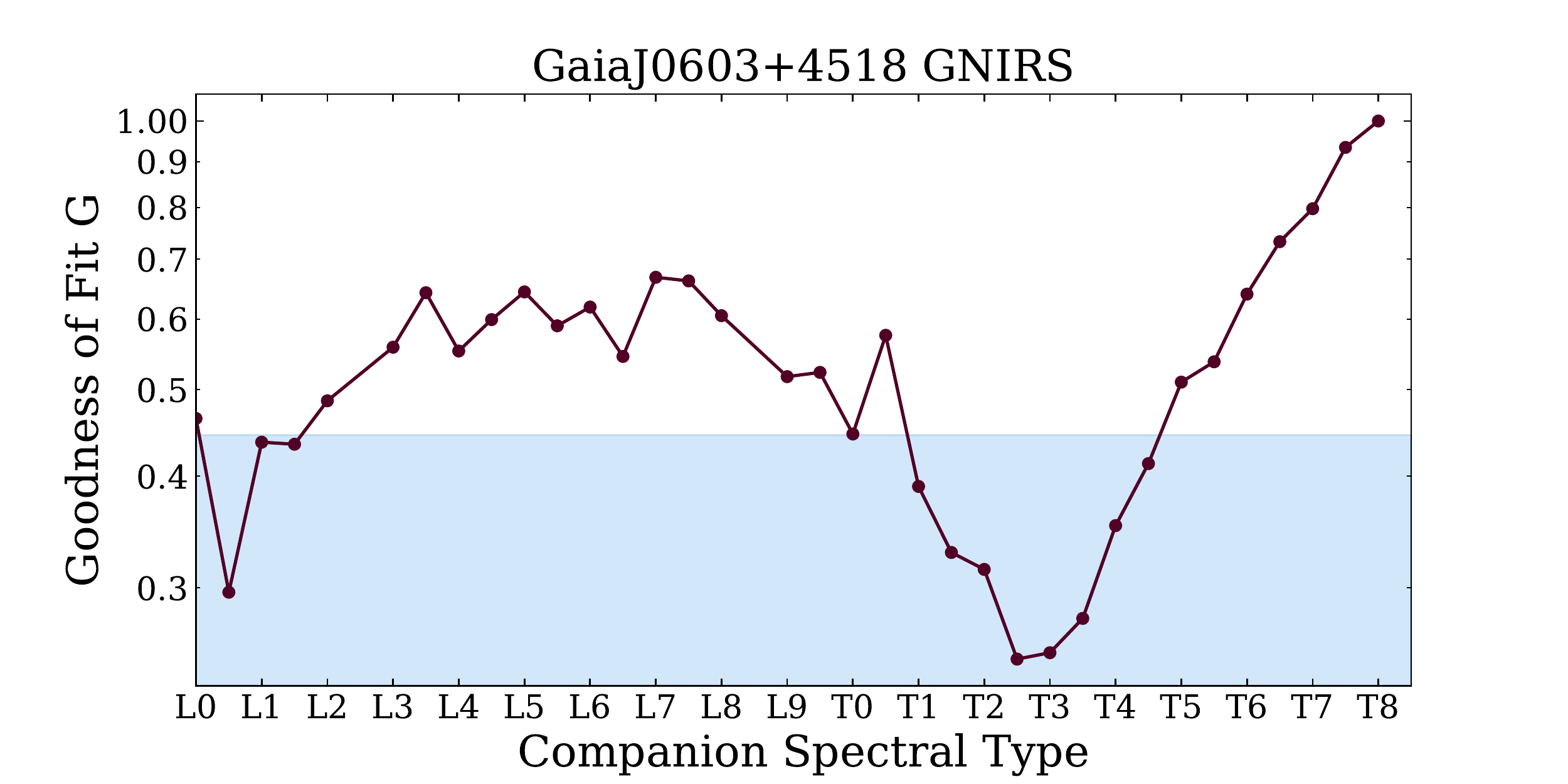}
    \includegraphics[width=0.65\textwidth]{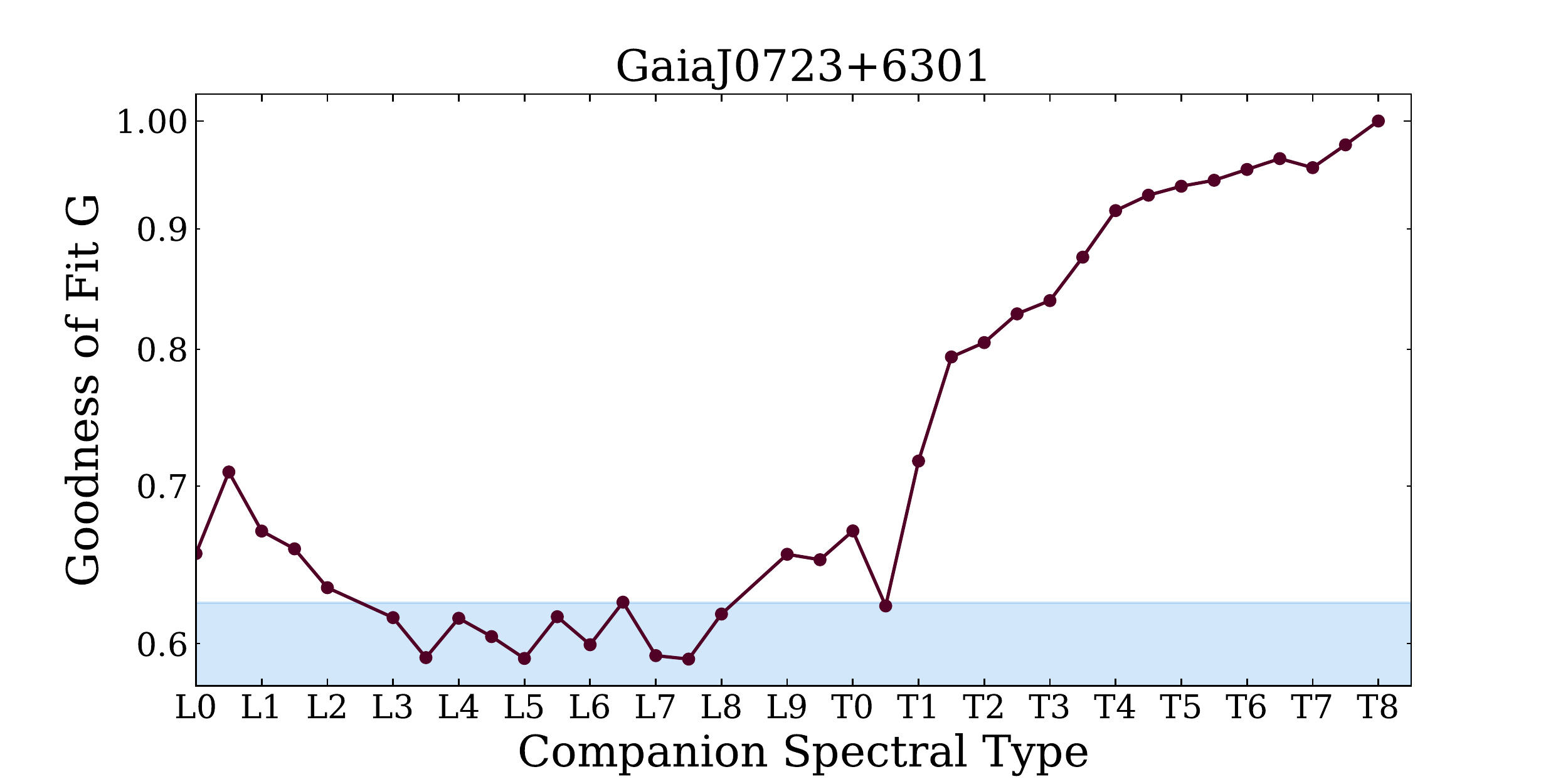}
    \caption{Brown dwarf companion goodness of fit statistic, \textit{G}, plotted against spectral types in the \textit{SpeX} library for the likely brown dwarf companion systems. The blue-shaded region shows the lower 32 percentile ($1 \sigma$) of average \textit{G} values for the spectral types. Note the different y-axis range for each system, as the Gaia~J0723+6301 fits were poor compared to those for Gaia~J0052+4505 and Gaia~J0603+4518.
    \label{fig: BD-G-plots}
    }
\end{figure*}

\begin{figure*}[tbh]
    \centering
    \includegraphics[width=0.65\textwidth]{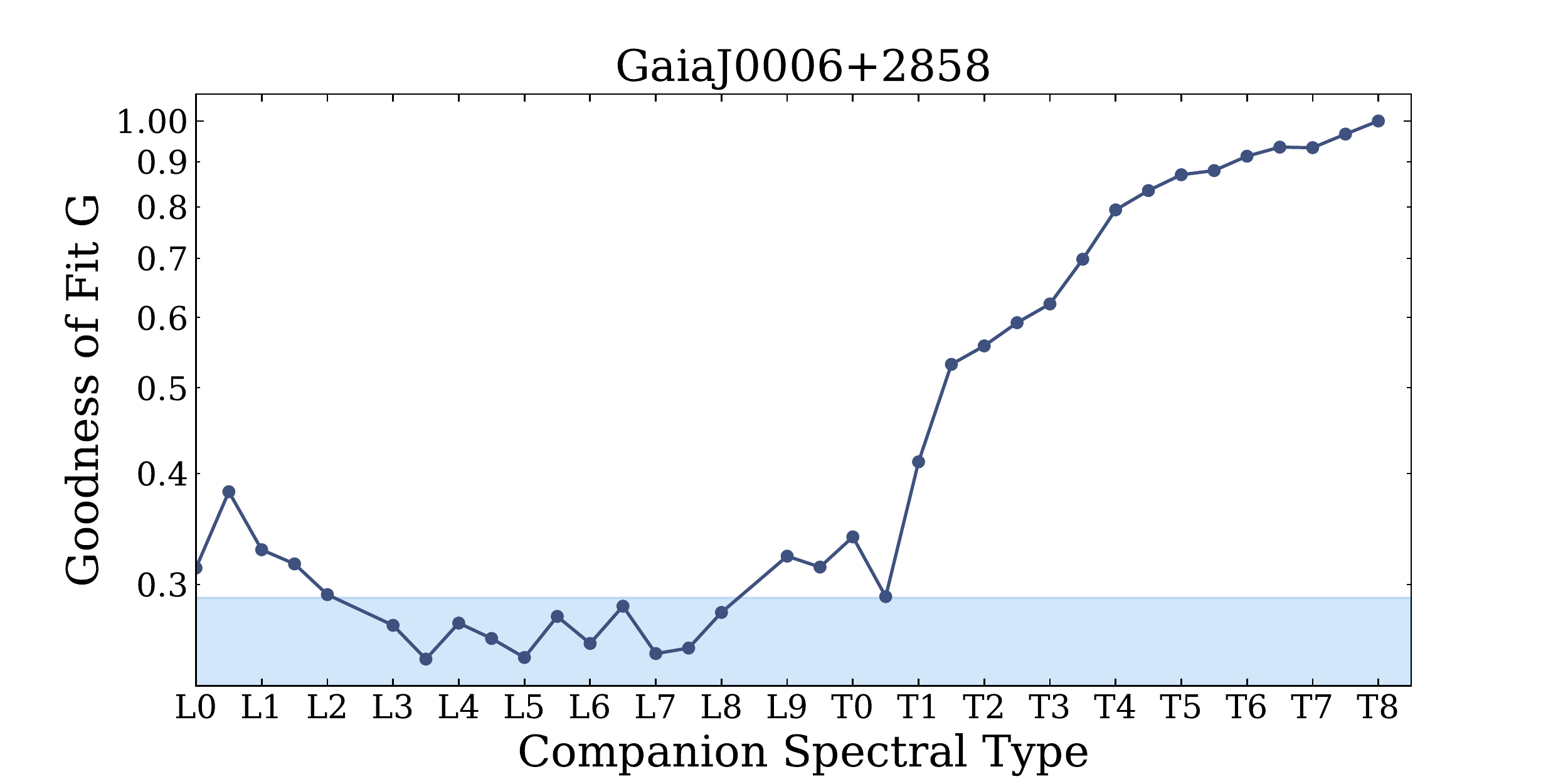}
    \includegraphics[width=0.65\textwidth]{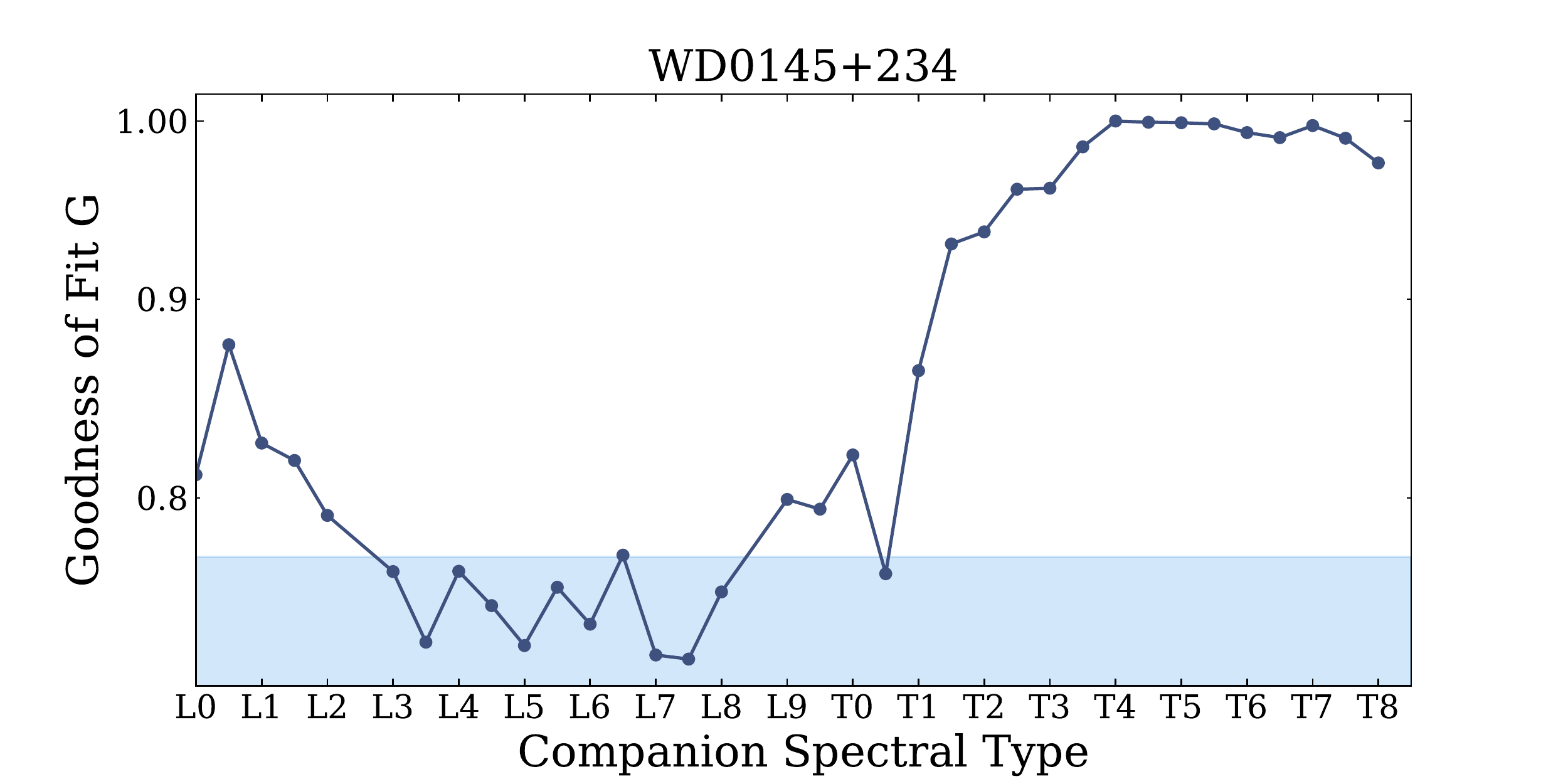}
    \includegraphics[width=0.65\textwidth]{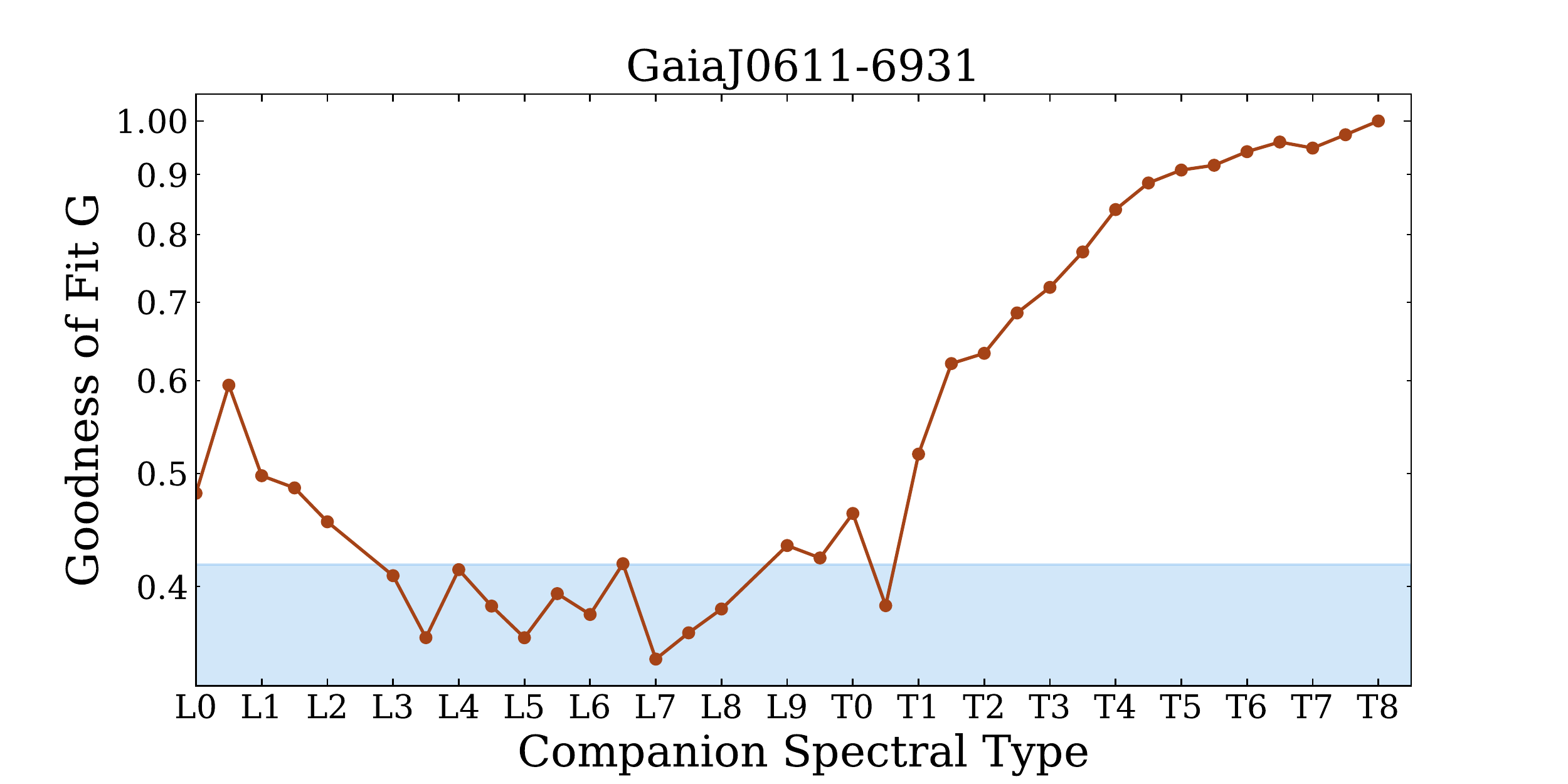}
    \includegraphics[width=0.65\textwidth]{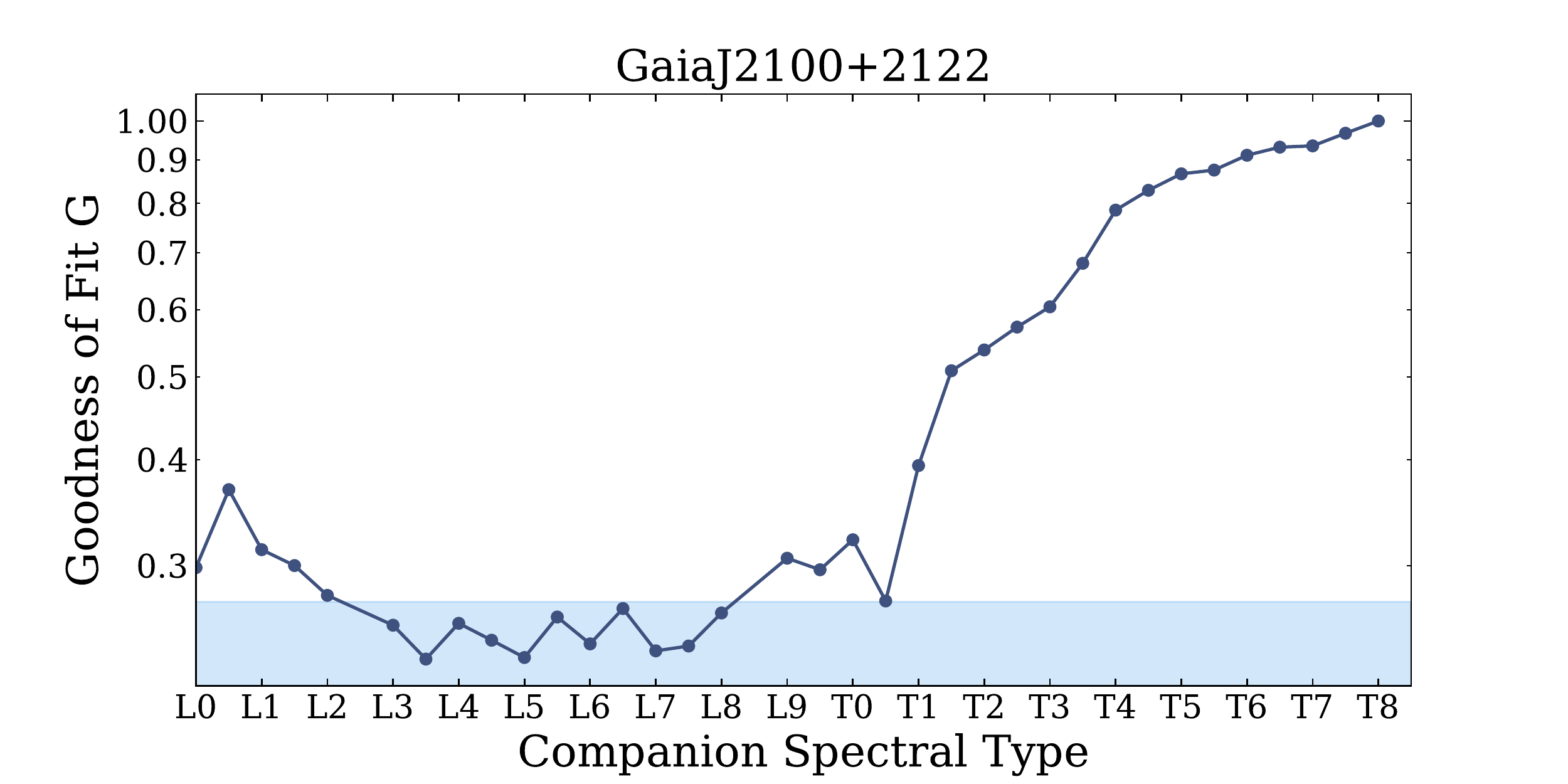}
    \caption{Similar to Figure~\ref{fig: BD-G-plots} except for the four disk systems.
    \label{fig: disk-G-plots}
    }
\end{figure*}

\clearpage

\bibliography{MOSFIRE-citations}
\bibliographystyle{aasjournal}

\end{document}